\documentclass[10pt,journal]{IEEEtran}

\usepackage{enumitem}
\usepackage{calc,amsfonts,amssymb,amsmath,bm,url,color,theorem,graphicx,cite,epstopdf,nicefrac,bbold}
\usepackage{psfrag,float}
\usepackage[algoruled,linesnumbered]{algorithm2e}
\usepackage{multirow}
\usepackage{dsfont}
\usepackage[normalem]{ulem}
\usepackage{stmaryrd}
\usepackage{xcolor}
\usepackage{psfrag,framed,mdframed}
\usepackage{comment}
\usepackage{mdframed}
\usepackage{threeparttable}
\usepackage{steinmetz}

\usepackage{amsmath}
\usepackage{graphicx}

\newcommand{\bc}{\boldsymbol{c}}

\newcommand{\g}{\boldsymbol{g}}

\renewcommand{\r}{\boldsymbol{r}}

\renewcommand{\v}{\boldsymbol{v}}
\newcommand{\w}{\boldsymbol{w}}
\newcommand{\x}{\boldsymbol{x}}

\newcommand{\z}{\boldsymbol{z}}

\newcommand{\zero}{\boldsymbol{0}}
\newcommand{\one}{\boldsymbol{1}}

\newcommand{\A}{\boldsymbol{A}}
\newcommand{\B}{\boldsymbol{B}}
\newcommand{\C}{\boldsymbol{C}}

\newcommand{\M}{\boldsymbol{M}}

\renewcommand{\S}{\boldsymbol{S}}

\newcommand{\U}{\boldsymbol{U}}

\newcommand{\X}{\boldsymbol{X}}
\newcommand{\Y}{\boldsymbol{Y}}
\newcommand{\Z}{\boldsymbol{Z}}

\newcommand{\argmin}[1]{\underset{#1}{\rm \text{arg min }}}

\newcommand{\btheta}{ {\boldsymbol{\theta}} }

\newcommand{\bOmega}{{ \bm \varOmega } }

\newcommand{\tM}{\underline{ \boldsymbol{M} }}

\newcommand{\tV}{\underline{ \boldsymbol{V} }}

\newcommand{\tX}{\underline{ \boldsymbol{X} }}
\newcommand{\tY}{\underline{ \boldsymbol{Y} }}

\newcommand{\gtM}{\underline{\boldsymbol{M} }^\natural}
\newcommand{\wtM}{\widetilde{\underline{\boldsymbol{M}} }}
\newcommand{\gtX}{\underline{\boldsymbol{X} }^\natural}
\newcommand{\otM}{\underline{\boldsymbol{M} }^\star}

\newcommand{\cA}{\mathcal{A}}
\newcommand{\cB}{\mathcal{B}}

\newcommand{\cG}{\mathcal{G}}

\newcommand{\cL}{\mathcal{L}}
\newcommand{\cM}{\mathcal{M}}
\newcommand{\cN}{\mathcal{N}}
\newcommand{\cO}{\mathcal{O}}

\newcommand{\cQ}{\mathcal{Q}}
\newcommand{\cR}{\mathcal{R}}
\newcommand{\cS}{\mathcal{S}}
\newcommand{\cT}{\mathcal{T}}

\newcommand{\cW}{\mathcal{W}}
\newcommand{\cX}{\mathcal{X}}

\newcommand{\bbR}{ {\mathbb{R}} }
\newcommand{\bbE}{ {\mathbb{E}} }

\newcommand{\bbP}{ {\mathbb{P}} }

\usepackage{textcomp}
\newcommand{\tpm}{\textpm }

\makeatletter
\newcommand{\vast}{\bBigg@{4}}
\newcommand{\Vast}{\bBigg@{5}}
\makeatother

\DeclareMathOperator*{\minimize}{\textrm{minimize}}

\usepackage{xcolor}
\usepackage{psfrag,framed}
\usepackage{lipsum}
\PassOptionsToPackage{normalem}{ulem}
\usepackage{ulem}

\definecolor{shadecolor}{RGB}{220,220,220}
\usepackage{caption}
\usepackage{subcaption}

\definecolor{orange}{RGB}{255,107,0}

\definecolor{green}{RGB}{51,150,30}

\newcommand{\ualpha}{ {\sf U_\alpha} }
\newcommand{\lalpha}{ {\sf L_\alpha} }
\newcommand{\falpha}{ {\sf F_\alpha} }

\newtheorem{Fact}{Fact}
\newtheorem{Lemma}{Lemma}

\newtheorem{Theorem}{Theorem}
\newtheorem{Def}{Definition}

\theorembodyfont{\rmfamily}

\newtheorem{Assumption}{Assumption}
\newtheorem{Remark}{Remark}

\begin{document}

\title{Quantized Radio Map Estimation Using Tensor and Deep Generative Models}

\author{Subash Timilsina, Sagar Shrestha, and Xiao Fu
 
\thanks{The authors are with the School of EECS at Oregon State University.
This work was supported in part by the National Science Foundation (NSF) under Projects NSF ECCS-2024058 and NSF CCF-2210004. 
}
}

\maketitle

\begin{abstract}
\emph{Spectrum cartography} (SC), also known as  \emph{radio map estimation} (RME), aims at crafting multi-domain (e.g., frequency and space) radio power propagation maps from limited sensor measurements. 
While early methods often lacked theoretical support, recent works have demonstrated that radio maps can be provably recovered using low-dimensional models---such as the {\it block-term tensor decomposition} (BTD) model and certain {\it deep generative models} (DGMs)---of the high-dimensional multi-domain radio signals.
However, these existing provable SC approaches assume that sensors send real-valued (full-resolution) measurements to the fusion center, which is unrealistic. 
This work puts forth a {\it quantized} SC framework that generalizes the BTD and DGM-based SC to scenarios where heavily quantized sensor measurements are used. A maximum likelihood estimation (MLE)-based SC framework under a Gaussian quantizer is proposed. 
Recoverability of the radio map using the MLE criterion is characterized under realistic conditions, e.g., imperfect radio map modeling and noisy measurements. 
Simulations and real-data experiments are used to showcase the effectiveness of the proposed approach. 

\end{abstract}

\begin{IEEEkeywords}
Radio map estimation, spectrum cartography, block-term tensor decomposition, deep generative model, Gaussian quantization.
\end{IEEEkeywords}

\section{Introduction}
The {\it spectrum cartography} (SC) [also known as {\it radio map estimation} (RME)] technique was proposed to build multi-domain (e.g., space, frequency, and time) radio maps from limited sensor measurements that are sparsely acquired over a geographical area; see, e.g., \cite{mateos2009spline, bazerque2011group, boccolini2012wireless, kim2013cognitive, jayawickrama2013improved, zhang2020spectrum, shrestha2022deep,teganya2020data}. The radio maps capture key characteristics of the radio frequency (RF) environment (e.g., interference propagation), and thus are critical for various tasks, such as opportunistic access, spectrum surveillance, beamforming, power allocation, and interference management; see \cite{bi2019engineering}.
From a signal processing viewpoint, estimating a multi-domain, high-dimensional and high-resolution radio map from limited sensor-acquired samples/measurements is an ill-posed inverse problem---where an infinite number of solutions exist in general, hindering the recoverability of the radio map.

Early SC/RME methods often assume that power propagation is smooth over the space. This made it possible to use various interpolation techniques for SC, e.g., the Kriging interpolation \cite{boccolini2012wireless}, thin plate splines \cite{bazerque2011group}, kernel methods \cite{romero2017learning}, and Gaussian radial basis functions (RBF) \cite{hamid2017non}. Sparse representations of the radio maps in certain domains are also often leveraged for SC; see, e.g., \cite{bazerque2009distributed, jayawickrama2013improved, kim2013cognitive,bazerque2011group}. 
In recent years,
low-rank matrix and tensor completion techniques were advocated for SC; see, e.g., \cite{khalfi2018airmap, schaufele2019tensor, zhang2020spectrum, sun2022propagation}.
From a low-rank model completion viewpoint, some of these methods were shown to ensure the recovery of the radio map; see, e.g., \cite{zhang2020spectrum,sun2022propagation}.
However, handcrafted priors such as sparsity and low rank do not always match with the reality, especially when heavy shadowing exists, e.g., in crowded urban or indoor environments; see discussions in \cite{shrestha2022deep}.

To better model heavily shadowed environments, a number of deep learning-based methods were employed for SC. 
Compared to handcrafted prior-based approaches, deep learning-based methods need a training phase using oftentimes a large amount of training data, which creates extra workload.
However, deep neural networks can represent complex scenarios as nonlinear functions in a succinct way, and thus can recover the radio map accurately under heavy shadowing. Earlier deep learning-based SC works in \cite{niu2018recnet, ratnam2020fadenet, han2020power, teganya2020data, krijestorac2021spatial} formulated the SC problem as image inpainting problems. 
The more recent work in \cite{shrestha2022deep} 
``embeds'' a {\it deep generative model} (DGM)-learned prior into the {\it spatial loss fields} (SLFs) of the emitters.
Then, a DGM-based data recovery problem was formulated. Compared to the inpainting-based methods, e.g., \cite{teganya2020data,han2020power}, the embedded DGM-based method in \cite{shrestha2022deep} enjoys a lighter training burden and exhibits better generalizability. 
More importantly, it was shown to ensure recoverability of the radio map under realistic conditions---and such theoretical guarantees had been lacking in prior deep learning-based SC works.

\noindent
{\bf Challenges.}
The recent developments of theory-backed SC (e.g., the works based on tensor \cite{zhang2020spectrum} and DGM \cite{shrestha2022deep}) were built upon the premise that the sensors communicate with the fusion center using full-resolution real-valued measurements. This setup is unrealistic as real-world communication systems often designate limited bandwidth for signaling. The measurements sent to the fusion center by the sensors are usually heavily quantized. 
Nonetheless, SC using quantized measurements was much less studied in the literature. 
The line of ``frugal sensing'' work \cite{mehanna2013frugal, konar2014parametric} considered 1-bit power spectrum estimation, but did not consider the reconstruction of the spatial information. The work in \cite{romero2017learning} used kernel regression for SC with quantized sensor feedback. However, it assumed that the power spectral density (PSD) of each emitter is known to the fusion center---but estimating the PSDs itself is a highly non-trivial task \cite{fu2015factor, fu2016power}. 
To our best knowledge, {\it blind} SC methods (i.e., SC without assuming knowing the emitters' PSDs) using quantized measurements and the associated theoretical understanding have not been studied---for both handcrafted prior and DGM-based SC frameworks.

\noindent {\bf{Contributions.}} In this work, we propose a quantized SC framework with provable recovery guarantees of the ground-truth radio map.
Our detailed contributions are as follows:

\noindent
$\bullet$ {\bf A Quantized SC Framework.} We propose a quantized SC framework that can flexibly work with various radio map models. We consider the setting where the measurements sent by the sensors are outputs of a Gaussian quantizer. 
Then, we formulate the quantized SC problem using the maximum-likelihood estimation (MLE) principle. We combine the quantization strategy with two radio map models, namely, {\it block-term tensor decomposition} (BTD) from \cite{zhang2020spectrum} and DGM from \cite{shrestha2022deep}, and design algorithms for tackling their respective MLE formulations.
Both models are well-motivated and useful: The former is more economical to deploy (as no training data is needed) and the latter is more resilient to heavy shadowing.

\noindent
$\bullet$ {\bf Recoverability Analysis.}
On the theory side,
we show that the formulated MLE criteria under both the tensor model and the DGM of the radio maps ensure recovering the ground truth (up to bounded errors), if reasonable conditions are met.
In particular, we show that our framework enjoys provable recovery even under challenging scenarios, e.g., when the low-rank tensor model or the DGM do not exactly match the ground truth.
The results also reveal an intuitive trade-off between the model complexity and the sample complexity.
Our analysis is a nontrivial integration of generalization error analysis and quantized data recovery, with careful consideration given to sensing paradigms in the context of SC.

\medskip

Part of the work will appear at ICASSP 2023 \cite{timilsina2023deep}, which introduced the basic idea and the DGM-based formulation. The journal version additionally includes 1) a tensor decomposition-based model that does not require off-line training; 2) detailed and unified recoverability analysis under both the DGM and BTD models; 3) more comprehensive simulations; and 4) real-data experiments.

\medskip

\noindent
\textbf{Notation} We use $x \in \bbR, \x \in \bbR^n, \X \in \bbR^{I \times J}, \tX \in \bbR^{I \times J \times K}$ to denote a scalar, vector, matrix, and tensor respectively. We adopt the Matlab notation $\X(i,:)$ and $\X(:,j)$ to represent the $i$th row and $j$th column of matrix $X$, respectively. $\tX(i,j,:)$ represents the $(i,j)$th tensor fiber. $\tX(i,j,k)$ represents the element of $\tX$ indexed by $i,j,k$. $\circ$ and $\circledast$ represent the outer product and the Hadamard product, respectively. The outer product between matrix $\U$ and vector $\v$ is defined as $\tX = \U \circ \v$ such that, $[\U\circ \v]_{i,j,k}=\U(i,j)\v(k)$.
$\zero$ and $\one$ represent the all-zero and all-one vectors, respectively. $\underline{\zero}$ and $\underline{\one}$ represent the all-zero and all-one tensors, respectively. $\| \x \|_2$ and $\| \X \|_2$ represent the $\ell_2$ norm and spectral norm, respectively. $\| . \|_{\rm F}$ denotes the Frobenius norm of matrices and tensors. We define $\| \tX \|_{\infty} = \max_{i,j,k}|\tX(i,j,k)|$.  $|\cX|$ denotes the cardinality of the set $\cX$. For an integer $I$, the set $[I]$ denotes $\{1,\ldots,I\}$. $\underline{\bm L}=\log(\tX)$ is a tensor such that $\underline{\bm L}({i,j,k})=\log( \tX({i,j,k}))$. 
``$\bm \geq $'' represents element-wise ``larger than or equal to'' (e.g., $\X\geq \bm 0$ means every element of $\X$ is nonnegative).
$\dot f(x)$ is the first-order derivative of $f$ at $x$.

\section{Background}

\begin{figure}[!t]
    \centering
    \begin{subfigure}{0.4\linewidth}
    \includegraphics[width=\linewidth]{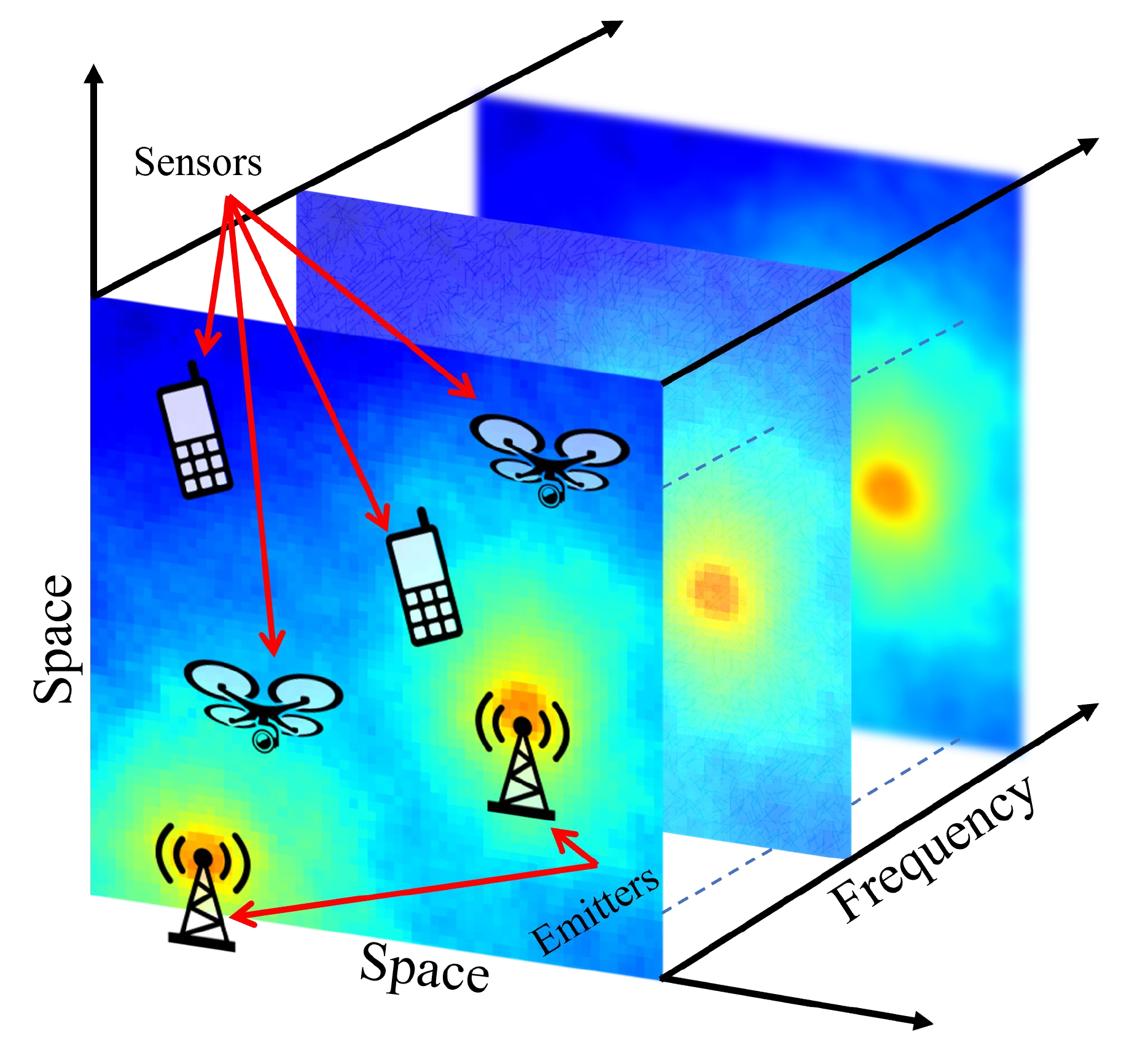}
    \caption{}
    \end{subfigure}
    \quad
    \begin{subfigure}{0.7\linewidth}
    \includegraphics[width=\linewidth]{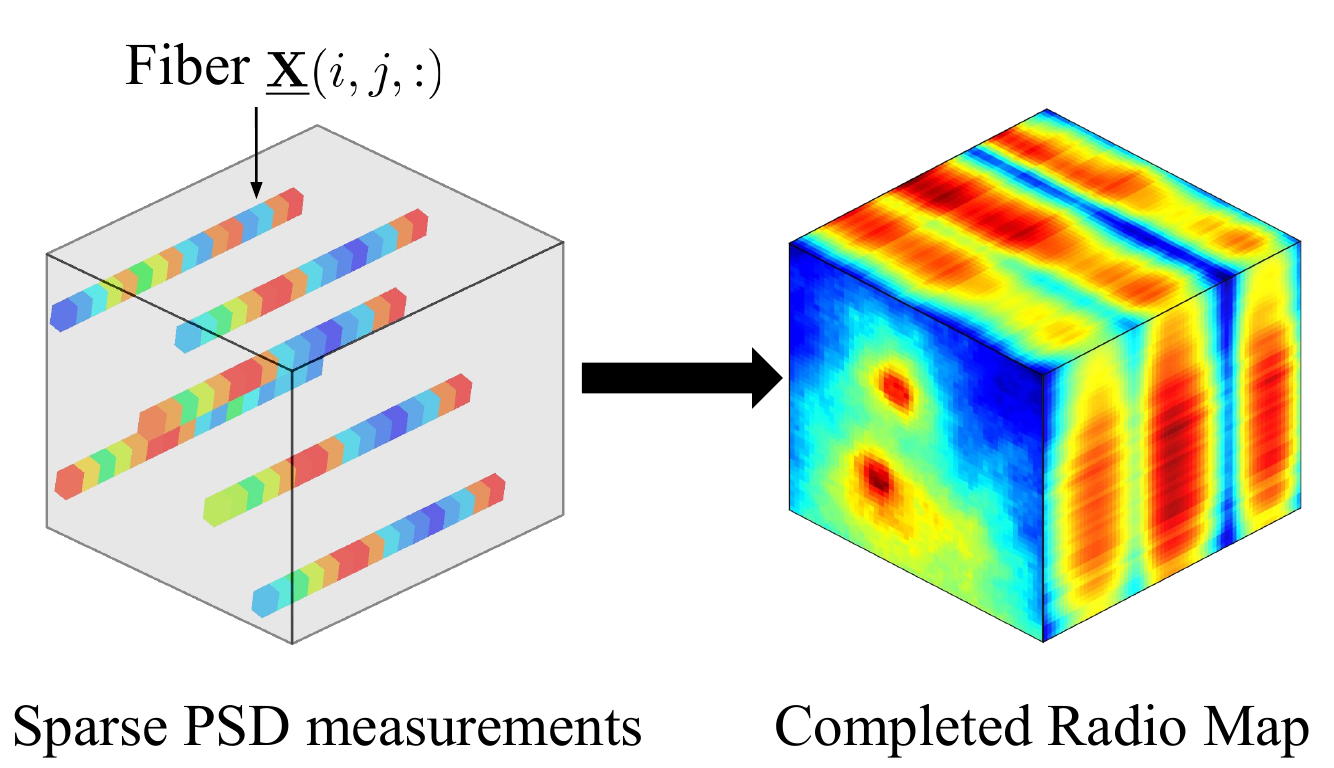}
    \caption{}
    \end{subfigure}
    \caption{(a) Scenario of interest. (b) Sampling pattern of the radio map; each sampled fiber is a PSD.}
    \label{fig:sc_illustration}
\end{figure}

\subsection{Problem Setup}\label{section:problem_setup}
We focus on the SC scenario illustrated in Fig.~\ref{fig:sc_illustration} (a). The scenario is widely considered in the literature; see, e.g., \cite{bazerque2011group, zhang2020spectrum, shrestha2022deep, romero2017learning}.
Specifically, we aim at recovering a {\it spatio-spectral} radio power density map induced by $R$ emitters using measurements acquired by $N$ sensors. The sensors are sparsely deployed over the region of interest.
For the ease of exposure, we assume that the space domain is a 2D rectangle and is discretized into $I \times J$ grids (but the proposed design principles can be readily generalized to 3D cases).
The frequency domain is also discretized into $K$ frequency bins---and thus
every spatial grid is associated with a $K$-dimensional {\it power spectral density} (PSD).
In other words, the power propagation patterns of the $R$ emitters over the $K$ frequencies constitute an $I\times J\times K$ radio map tensor $\tX \in \bbR^{I \times J \times K}$, where the entry $\tX(i,j,k)$ is the PSD of the signal received at location $(i, j)$ and frequency $k$. 
That is, every {\it fiber} \cite{fu2020block} of the tensor, $\tX(i,j,:)$, represents the PSD of the received signal measured at the location $(i, j)$; see Fig. \ref{fig:sc_illustration} (b). We use the notation $$\bOmega = \{ (i,j) | i \in [I], j \in [J] \}$$ to denote the set of sensor locations. Note that we often have $$|\bOmega| = N \ll IJ;$$
i.e., only a small number of sensors are available.
We assume that every sensor acquires the full PSD $\tX(i_s,j_s,:)$ at its location $(i_s,j_s)$, where $s=1,\ldots,N$.
If the sensors are able to transmit real-valued feedback to the fusion center,
the goal of SC is to recover the full $\tX$ from the tensor fibers $\{ \tX(i,j,:)  \}_{(i,j)\in\bOmega }$ at the fusion center.

\subsection{Prior Works on Provable SC}
Early developments of SC (see, e.g., \cite{boccolini2012wireless,bazerque2009distributed,bazerque2011group,kim2013cognitive,khalfi2018airmap,schaufele2019tensor}) mostly focused on the methodology side but less considered the theoretical aspects, e.g., recoverability of the ground-truth radio map. 
More recently, there has been an increased research interest on theoretical understanding to SC.
In this subsection, we briefly review two SC models that were shown to guarantee recovery of the radio map.

\subsubsection{Block-term Tensor Modeling and Provable Recovery} 
The recent work \cite{zhang2020spectrum} proposed a recoverability-guaranteed SC method from a block-term tensor completion viewpoint.
\begin{figure}[t!]
    \centering
    \includegraphics[width=0.8\linewidth]{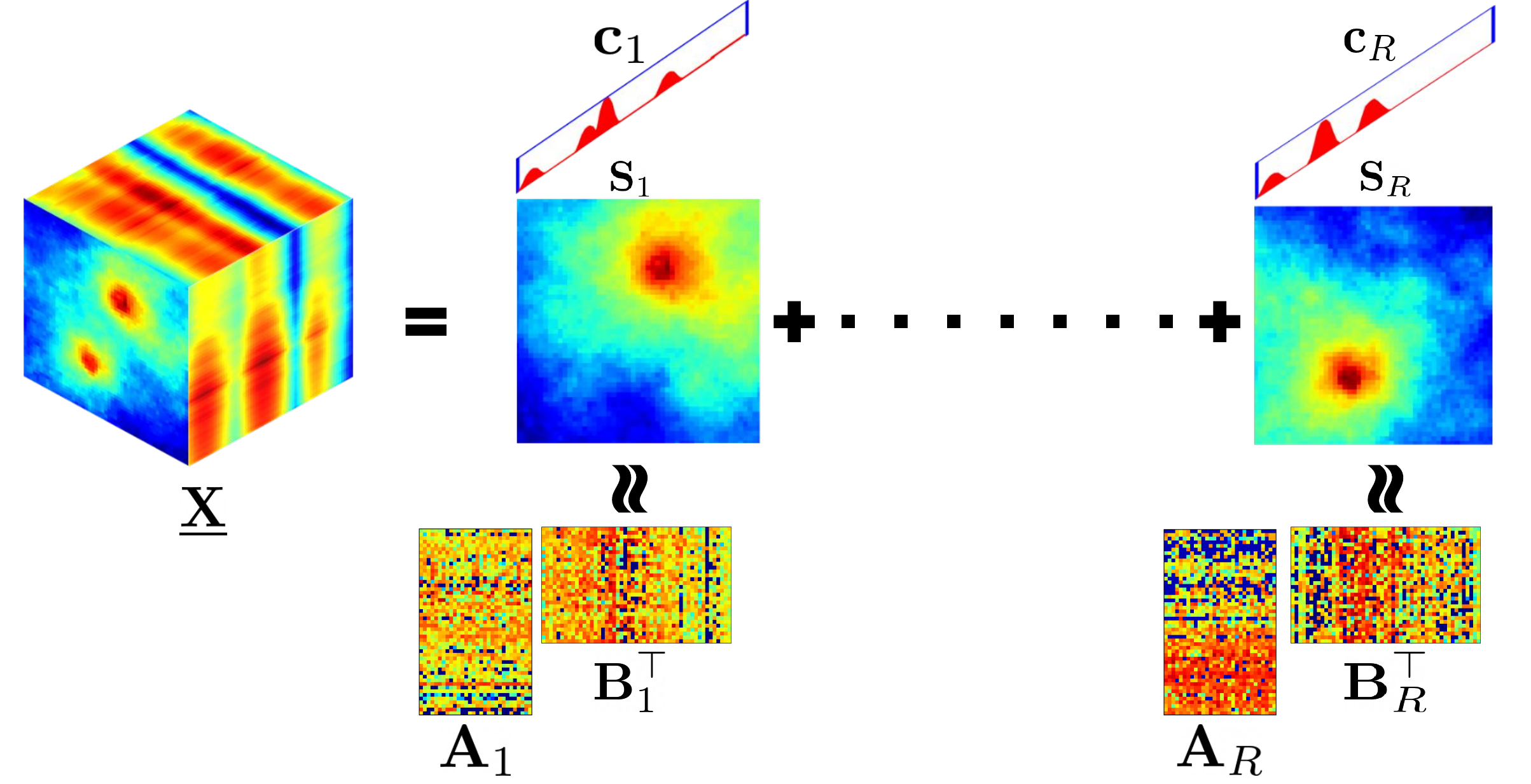}
    \caption{The BTD-based model for radio maps used in \cite{zhang2020spectrum}.}
    \label{fig:lnfacBTD}
\end{figure}
The model in \cite{zhang2020spectrum} starts by assuming that power propagation is coherent over the frequency band of interest, which in general holds when the ratio between the bandwidth of the frequencies of interest and its central frequency is not large \cite{polo2009compressive, bazerque2009distributed, fu2015factor, fu2016power}. Under such circumstances, the radio map can be decomposed into the latent factors associated with spatial and spectral information, respectively \cite{zhang2020spectrum, shrestha2022deep, bazerque2011group, romero2017learning}:
\begin{align}
    \tX(i,j,k) = \sum_{r=1}^R \S_r(i,j)\bc_r(k)\Longleftrightarrow \tX=  \sum_{r=1}^R \S_r \circ \bc_r,
    \label{eq:emitter_disaggregated}
\end{align}
where $\S_r \in \bbR^{I \times J}$ is the {\it spatial loss field} (SLF) of emitter $r$, $\bm c_r \in \bbR^K$ is the PSD of emitter $r$, and $\circ$ denotes the outer product. The SLF captures the spatial power propagation characteristics of an emitter, and the PSD reflects the emitter's spectral band occupancy.

The work \cite{zhang2020spectrum} modeled each SLF $\S_r$ in \eqref{eq:emitter_disaggregated} as a low-rank matrix, i.e., $$\S_r=\A_r\B_r^\top,$$ 
where $\A_r\in\mathbb{R}^{I\times L}$ and $\bm B_r\in\mathbb{R}^{J\times L}$ with ${\rm rank}(\A_r)={\rm rank}(\B_r)=L\ll \min\{I,J\}$. 
Fig.~\ref{fig:lnfacBTD} illustrates the model.
This low-rank model of $\S_r$ connected \eqref{eq:emitter_disaggregated} with a tensor model, namely, the {\it block-term tensor decomposition} (BTD) with multilinear rank-$(L,L,1)$ model\cite{de2008decompositions}.

The model is well-motivated, as the individual SLFs often exhibit high correlations across the spatial domain, making $\S_r$'s approximately low-rank; see \cite{zhang2020spectrum} for numerical evidence.

Using the BTD model of the radio map, the SC problem is formulated as follows:
\begin{align}    \label{eq:btd}
    \minimize_{\{\bc_r, \A_r,\B_r\}_{r=1}^R} \left \| \tM_{\text{sens}} \circledast \left ( \tX -  \sum_{r=1}^R (\A_r\B_r^\top) \circ \bc_r   \right) \right \|_{\rm F}^2,
\end{align}
where $\tM_{\text{sens}} $ is a sensing mask tensor such that $\tM_{\text{sens}}(i,j,:) = \one \in \mathbb{R}^K$ if $(i,j) \in \bOmega$ and $\tM_{\text{sens}}(i,j,:) = \zero$ otherwise. 
Notably, \cite{zhang2020spectrum} leveraged the BTD model's essential uniqueness to show that the spatio-spectral radio map is provably recoverable under both random and {\it regular} sampling patterns.

\subsubsection{Deep Generative Model (DGM)-Based Provable SC}
\begin{figure}[t!]
    \centering
    \includegraphics[width=0.8\linewidth]{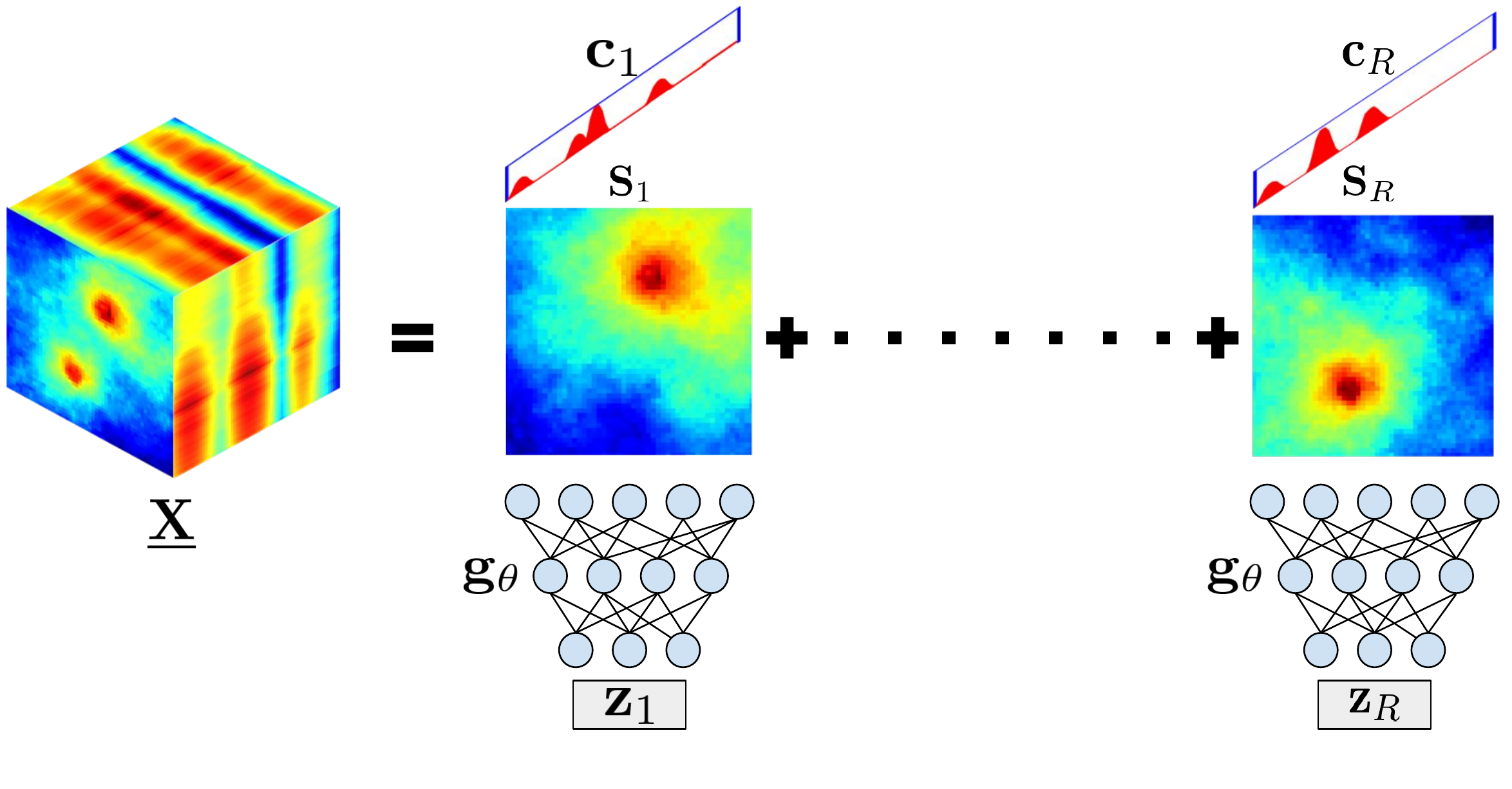}
    \caption{Spectrum cartography using deep generative model from \cite{shrestha2022deep}.}
    \label{fig:lnfac}
\end{figure}
In recent years, deep learning-based SC attracted much attention. The reason is that deep neural networks can represent very complex and heavily shadowed scenarios in a parsimonious way, and thus can boost performance of SC in challenging scenarios.
However, most early attempts, e.g., those in \cite{teganya2020data, han2020power}, formulated the SC problem as a supervised inpainting network learning problem. This formulation encountered training and generalization challenges, as the radio maps often have an extremely large latent state space. In addition, there is a lack of theoretical support to this line of work.

\begin{figure}[t!]
    \centering
    \includegraphics[width=0.8\linewidth]{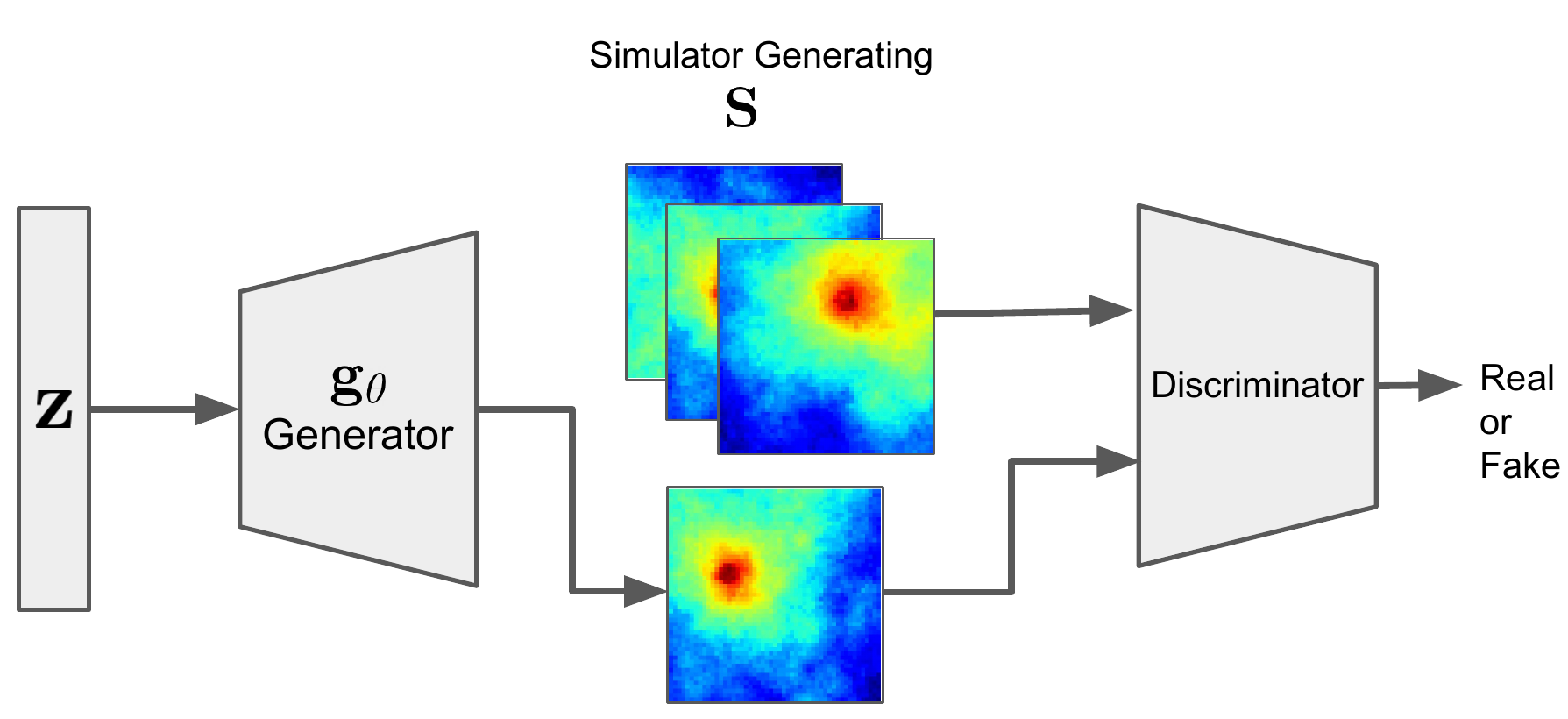}
    \caption{Using simulated SLFs to learn a DGM of the SLFs via GAN.}
    \label{fig:gan_train}
\end{figure}

To exploit the representation power of deep neural networks with recoverability guarantees, 
the work \cite{shrestha2022deep} provided an alternative solution.
The method can be regarded as an extension of the model in \cite{zhang2020spectrum}.
Realizing that the low-rank model for the SLFs in \cite{zhang2020spectrum} may not always hold in practice, \cite{shrestha2022deep} modeled the SLF of every emitter using a data-driven DGM. To be specific, \cite{shrestha2022deep} proposed to learn a DGM of the SLF by simulating a large number of SLFs following a physical model. Then, every $\S_r$ can be approximated as follows:
\begin{align}\label{eq:generative}
    \S_r \approx \g_{\btheta}(\z_r),
\end{align}
where $\g_{\btheta}(\cdot): \bbR^{D} \to \bbR^{I \times J}$ is a deep generative neural network that maps the ``latent embedding'' $\z_r\in\mathbb{R}^D$ such that $D \ll IJ$ to an SLF---see Fig.~\ref{fig:lnfac}. 
The DGM is learned off-line, e.g., via autoencoder, variational autoencoder (VAE), or generative adversarial network (GAN); see Fig. \ref{fig:gan_train}.
Using this learned DGM $\g_{\btheta}(\z_r)$, the SC problem was cast as follows:
\begin{align}
    \minimize_{\{\bc_r, \z_r\}_{r=1}^R} \left \| \tM_{\text{sens}} \circledast \left ( \tX -  \sum_{r=1}^R \g_{\btheta}(\z_r) \circ \bc_r   \right) \right \|_{\rm F}^2.
    \label{eq:leastsquares_baseline}
\end{align}
The method used a DGM to describe each {\it individual} SLF (instead of the entire radio map as in \cite{han2020power, zhang2020spectrum}). Consequently, the DGM is much easier to learn in the training stage compared to the networks in \cite{teganya2020data, han2020power}, which used a neural networks to learn the {\it aggregated} radio map from multiple emitters. This is because each SLF has a much smaller ``state space'' compared to that of the aggregated radio map; see more discussions in \cite{shrestha2022deep}.
The work in \cite{shrestha2022deep} also showed that \eqref{eq:leastsquares_baseline} guarantees recovering the radio map under reasonable conditions.

\begin{Remark}\label{remark:btdvsdgm}
    The BTD and DGM based methods have their respective advantages and disadvantages. The BTD model is simple to implement, but
    the {\it key limitation} lies in the designated low-rank prior for the SLFs. The low rank of $\S_r$ could be violated when the spatial correlation becomes weak---which typically happens when the environment has heavy shadowing effects, e.g., in urban areas. 
    The DGM-based approach is capable of ``encoding'' complex, non-analytical prior information using the neural representation of $\g_{\btheta}(\cdot)$, which is suitable for modeling heavily shadowed cases.
    In addition, the DGM parameterization may substantially reduce the number of unknown parameters (if ${\rm card}(\bm z_r )=D \ll IJ$), which can greatly reduce the sample complexity of SC.
However, the price to pay is that an extra offline training stage is needed. To generate training data for learning $\bm g_{\bm \theta}(\cdot)$, a certain level of awareness of the shadowing situation is required. These were not needed in the BTD work \cite{zhang2020spectrum}. Hence, both models are useful and meaningful. The choice of the model depends on the scenario of interest.
\end{Remark}

\subsection{Challenges - Lack of Quantized SC Approaches}
Most of the previous SC works, including the BTD and DGM based methods, assumed that real-valued measurements (e.g., $\tX(i,j,:)$ for all $(i,j)\in \bm \Omega$) can be transmitted to the fusion center (see, e.g., \cite{bazerque2011group, boccolini2012wireless, kim2013cognitive, jayawickrama2013improved, zhang2020spectrum, shrestha2022deep}). However, sending real-valued measurements is costly in terms of communication overhead. It is much more realistic that the sensors send quantized measurements. Quantized spectrum sensing was studied before (see, e.g., \cite{han2010efficient, lee2012spectrum} and the ``frugal sensing'' works \cite{mehanna2013frugal, konar2014parametric}) but the spatial information was not taken into consideration. Quantized SC was tackled in \cite{romero2017learning} using kernel regression. However, \cite{romero2017learning} assumed that the emitters' PSDs are known, but estimating the PSDs itself is a hard problem \cite{fu2015factor,fu2016power}.

In addition to the lack of effective methodology,
the theoretical aspects of quantized SC have yet to be studied. While the DGM-based framework in \cite{shrestha2022deep} and the BTD-based framework in \cite{zhang2020spectrum} provided provable recoverability of the radio map with real-valued measurements, it remains an open question how to extend these results to handle quantized measurements without losing recoverability supports.

\section{Proposed Approach}

In this work, we propose a {\it quantized SC} framework where the sensors only feedback heavily quantized measurements.
In particular, we will design a system that can flexibly work with the BTD and DGM models of the radio map in \cite{zhang2020spectrum} and \cite{shrestha2022deep}.

\subsection{Sensing and Quantization}
Following the setup in Sec. \ref{section:problem_setup}, we assume that the sensor located at $(i,j)$ can acquire the PSD of the received signal, i.e., $\tX(i,j,:)\in\mathbb{R}^K$. In addition, we consider the scenario where the sensors send a quantized version of $\tX(i,j,:)$. To be more specific, we employ a quantizer $${\cal Q}(\cdot):\mathbb{R}\rightarrow \mathbb{Z}$$ that maps each element of the PSD received at $(i,j)$ to an integer, i.e.,
\[      \tY(i,j,k) ={\cal Q}(\tX(i,j,k)),~k=1,\ldots,K  \]
where $\tY(i,j,:)\in\mathbb{Z}^K$ is the quantized PSD.

We employ the Gaussian quantization strategy that is widely in the literature \cite{davenport20141, bhaskar2016probabilistic, cao2015categorical, ghadermarzy2018learning, li2018tensor, mccullagh1980regression, lee2020tensor}.
To be specific, the quantization strategy is expressed as follows:
\begin{align}    \label{eq:model_qsc}
    &\tY(i,j,k) = \cQ(h( \tX(i,j,k) ) +\tV(i,j,k) )\\
    &\cQ(x) = q, \text{ if } b_{q-1} < x \leq b_{q}, \quad q \in [Q] , \nonumber
\end{align}
where $\tV(i,j,k) \sim \cN(0,\sigma^2)$ for all $i,j,k$ are i.i.d. zero-mean Gaussian noise, and $\{b_q\}_{q=1}^Q$ are the pre-specified quantization bins (i.e., $[b_{q-1},b_q]$ is the $q$th quantization interval). Here, the function $h(\cdot)$ is an invertible transformation function, 
\begin{align}\label{eq:logtrans}
h(x)=\log(x + a),     
\end{align}
where $a>0$. For notation simplicity, we slightly abuse the notation by applying $h$ to both entries and tensors, i.e., $h(\tX)$ means taking the transform in \eqref{eq:logtrans} for each entry of $\tX$.
Note that the function $h(\cdot)$ was not used in classic Gaussian quantization. In this work, we use this function to control the dynamic range of the data (also see Sec. \ref{section:quantization_design} for more discussions).
Note that as $h$ is invertible, it does not lose information of $\tX$; i.e., recovering $h(\tX)$ from the quantized data recovers $\tX$.

Quantization with artificial noise like in \eqref{eq:model_qsc} is called {\it dithering} in signal processing \cite{schuchman1964dither}. Dithering is known to be beneficial for retaining more information about the original signal in the quantized version by reducing the correlation between the quantization error and original signal \cite{lipshitz1992quantization}.

\subsection{Maximum Likelihood Estimation}
Let $\tM = h(\tX)$ be the transformed measurements of radio map. The quantized observations $\tY$ has the following distribution: 
\begin{align}
    \tY(i,j,k) = q,& \text{ with probability}~ f_q(\tM(i,j,k)), \nonumber \\ &\forall (i,j) \in \bOmega, k \in [K]
    \label{eq:quantized_observation}
\end{align}
where $f_q$ is defined as,
\begin{align}\label{eq:linkfunc}
    &f_q(\tM(i,j,k)) = \bbP(\tY(i,j,k) = q ~|~ \tM(i,j,k))\nonumber\\
    &~= \Phi(b_q - \tM(i,j,k)) - \Phi(b_{q-1} - \tM(i,j,k)),
\end{align}
in which $\Phi(x)$ is the cumulative distribution function (CDF) of the zero-mean and $\sigma^2$-variance Gaussian variable (i.e., the CDF of the dithering noise in \eqref{eq:model_qsc}).
Using \eqref{eq:linkfunc} and a certain parameterization of the radio map, one can formulate the SC problem as an MLE problem. In the next two subsections, we will use the BTD model in \cite{zhang2020spectrum} and the DGM model in \cite{shrestha2022deep} to formulate two MLE criteria.

\subsubsection{BTD-Based Quantized SC}\label{section:btd_formulation} As in \cite{zhang2020spectrum}, we first represent the radio map using the BTD model (see Fig.~\ref{fig:lnfacBTD}), i.e.,
$$ \tX^{\rm BTD} =\sum_{r=1}^R (\A_r\B_r^\top)\circ \bm c_r. $$
Using the described sensing and quantization procedure as in the previous section, the fusion center receives a quantized and incomplete tensor $\tY(\bm \varOmega,:)$.
To recover the ground-truth radio map, we consider the following MLE:
\begin{mdframed}
{\sf MLE via BTD}
\begin{align}\label{eq:btd_MLE}
    &\minimize_{(\A,\B,\C) \in  {\cal S}^{\rm BTD}}~F^{\rm BTD}_{\bm \varOmega,\tY}(\A,\B,\C), 
\end{align}
\end{mdframed}
where the objective function is defined as
\begin{align*}
&F^{\rm BTD}_{\bm \varOmega,\tY}(\A,\B,\C) =\\
 &\quad   -\sum_{(i,j) \in \bOmega} \sum_{k=1}^K \sum_{q=1}^Q \mathbb{1}_{[\tY(i,j,k) = q]}   \log ( f_{q}  \big(h ( \tX^{\rm BTD}(i,j,k) ) \big) ),
\end{align*}
in which we have
$ \tX^{\rm BTD}(i,j,k) = \left[ \sum_{r=1}^R (\A_r\B_r^\top ) \circ \bc_r  \right]_{i,j,k}$, $\A_r \in \bbR^{I \times L}, \B_r \in \bbR^{J \times L}$ and $\C = [\bc_1,\ldots,\bc_R]$. In addition, we have used the notation $\A=[\A_1,\ldots,\A_R]$ and $\B=[\B_1,\ldots,\B_R]$.
The constraint set is defined as follows:
\begin{align}\label{eq:sln_set_ABC}
   {\cal S}^{\rm BTD} & = \bigg \{ (\A,\B,\C)~|~ \A\geq\bm 0,\B\geq\bm 0,\C\geq\bm 0 \\
   &  \| \A_r\|_{\rm F}\leq \sqrt{\beta},\|\B_r \|_{\rm F} \leq \sqrt{\beta},  \| \bc_r \|_2 \leq \kappa, \forall r \in [R] \bigg \}. \nonumber
\end{align} 

 The constraints in \eqref{eq:sln_set_ABC} 
mean that the SLFs and the PSDs are both nonnegative and bounded (with $\|\S_r\|_{\rm F}\leq \beta$)---which are mild assumptiions per their physical meaning.
Particularly, we assume that the nonnegativity and boundedness of $\S_r=\A_r\B_r^\top$ are realized via having constraints on the latent factors of the BTD model. We also define the corresponding set:
\begin{align}\label{eq:sln_set_X_btd}
    &{\cal X}^{\rm BTD} =\\
    &\bigg\{\tX~\bigg|~\tX=\sum_{r=1}^R(\A_r\B_r^\top)\circ \bm c_r,~(\bm A,\B,\bm C)\in {\cal S}^{\rm BTD} \bigg \}. \nonumber
\end{align}

\subsubsection{DGM-Based Quantized SC}
As mentioned, the BTD model-based approach may not work well when the spatial region exhibits heavy shadowing \cite{shrestha2022deep}.
In such cases, it is natural to employ the DGM-based model in \eqref{eq:generative} to parameterize the radio map, i.e.,
\begin{equation}\label{eq:deep_radio_map}
    \tX^{\rm DGM} = \sum_{r=1}^R \bm g_{\bm \theta}(\z_r) \circ \bm c_r,
\end{equation}
where $\bm g_{\bm \theta}(\cdot)$ is learned in an off-line manner as in \cite{shrestha2022deep}; see Fig.~\ref{fig:lnfac}.
The radio map model in \eqref{eq:deep_radio_map} leads to the following MLE-based recovery criterion:
\begin{mdframed}
{\sf MLE via DGM}
\begin{align}\label{eq:MLE}
    \minimize_{(\C,\Z)\in {\cal S}^{\rm DGM}}~F_{\bm \varOmega,\tY}^{\rm DGM}(\bm Z,\bm C),   
\end{align}
\end{mdframed}
where  we have
\begin{align}
&F_{\bm \varOmega,\tY}^{\rm DGM}(\bm Z,\bm C)= \\
  &      -\sum_{(i,j) \in \bm \varOmega} ~\sum_{k=1}^K \sum_{q=1}^Q \mathbb{1}_{[{\tY(i,j,k) = q}]}   
    \log ( f_{q}  \big(h ( \tX^{\rm DGM}(i,j,k) ) \big) ), \nonumber
\end{align}
in which $  \tX^{\rm DGM}(i,j,k) = [ \sum_{r=1}^R \g_{\btheta}(\z_r) \circ \bc_r  ]_{i,j,k} $, $\C = [\bc_1,\ldots,\bc_R]$, and $\Z=[\z_1,\ldots,\z_R]$. 
The constraint set is defined as follows:
\begin{align}\label{eq:sln_set_CS}
    {\cal S}^{\rm DGM}& =  \bigg \{(\C,\Z) ~|~ \bm g_{\theta}(\z_r)\geq\bm 0,\bm C\geq \bm 0,  \\
&  \| \bm g_{\theta}(\z_r) \|_{\rm F} \leq \beta , \| \bc_r \|_2 \leq \kappa, \forall r \in [R] \bigg \}, \nonumber
\end{align}
where $\beta$ and $\kappa$ are positive constants.
Similar as before, we define a corresponding set in the data domain:
\begin{align}\label{eq:sln_set_X_dgm}
  &  {\cal X}^{\rm DGM} = \\
   & \bigg\{\tX~\bigg|~\tX=\sum_{r=1}^R\bm g_{\bm \theta}(\z_r)\circ \bm c_r,~(\bm Z,\bm C)\in {\cal S}^{\rm DGM} \bigg\}. \nonumber
 \end{align}

\medskip

As in the continuous measurement case (cf. Remark~\ref{remark:btdvsdgm}), the BTD-based formulation in \eqref{eq:btd_MLE} is better suited for cases where training data is unavailable or the training cost is not affordable.
The DGM-based formulation in \eqref{eq:MLE} has a more expressive model to attain enhanced recovery accuracy, especially under challenging scenarios, e.g., heavy shadowing.

\section{Optimization and Implementation}
In this section, we propose algorithms for handling the MLE problems in \eqref{eq:btd_MLE} and \eqref{eq:MLE}.
\subsection{Algorithm For BTD-Based MLE}
To tackle Problem~\eqref{eq:btd_MLE}, we consider the following reformulated approximation:
\begin{align}\label{eq:MLE_btd}
    \minimize_{\A,\B,\C}&~ \cL_{\bOmega, \tY}(\A, \B, \C) \\
    {\rm subject~to}&~ \A \geq \bm 0, \B \geq \bm 0, \C \geq \bm 0, \nonumber
\end{align}
where the objective function is expressed as follows: 
\begin{align*}\cL_{\bOmega, \tY}(\A, \B, \C) = & F^{\rm BTD}_{\bm \varOmega,\tY}(\A,\B,\C) + \lambda_1 r_1(\bm A) \\&+ \lambda_2 r_2(\bm B) +  \lambda_3 r_3(\bm C),
\end{align*} 
in which the regularization terms are as follows:
\[ r_i(\cdot)=\|\cdot\|_{\rm F}^2,~i=1,2,3.\]
Note that promoting small $\|\A\|_{\rm F}^2$ and $\|\B\|_{\rm F}^2$ encourages small-energy SLFs $\|\A_r\B_r^\top\|_{\rm F}$.
Hence, such regularization terms {\it indirectly} enforce the constraints in \eqref{eq:sln_set_ABC}, i.e., that the SLFs and the PSDs are bounded. 
Nonetheless, using the regularized reformulation instead of the constraints in \eqref{eq:sln_set_ABC} is more convenient to for designing optimization algorithms.

We propose a block coordinate descent (BCD) procedure to handle \eqref{eq:MLE_btd}.
In the $k$th iteration, we first update $\C$ as follows:
\begin{align}\label{eq:btd_c}
    \C^{(k+1)} \gets \max(& \C^{(k)} - \xi_3^{(k)}\overline{\nabla}_{\C} (  \cL_{\bOmega, \tY}(\A^{(k)}, \B^{(k)}, \C^{(k)}) \nonumber\\
    &+ \lambda_3 r_3 (\bm C) ),\bm 0),
\end{align}
where
$\overline{\nabla}_{\C}(\cdot)$ is a gradient-related direction of $F_{\bOmega, \tY}$ w.r.t $\C$.
Using the gradient w.r.t. $\C$\footnote{ Throughout this paper, the gradient computations are all implemented numerically by \texttt{PyTorch} \cite{autograd}. }, the direction $\overline{\nabla}_{\C}(\cdot)$ is found by popular methods such as plain-vanilla gradient and momentum-assisted methods like \texttt{Adagrad} \cite{duchi2011adaptive} and \texttt{Adam} \cite{kingma2014adam}), and $\xi_3^{(k)}$ is the step size. The ``$\max(\cdot,0)$'' step projects the updated $\C^{(k)}$ back to the nonnegative orthant.

Similarly, the update of $\A$ and $\B$ can be done as follows:
\begin{align}
    \A^{(k+1)} \gets \max(& \A^{(k)} - \xi_1^{(k)}\overline{\nabla}_{\A} (  \cL_{\bOmega, \tY}(\A^{(k)}, \B^{(k)}, \C^{(k)}) \nonumber \\
    &+ \lambda_1 r_1 (\bm A) ),\bm 0), \label{eq:btd_a}\\
    \B^{(k+1)} \gets \max(& \B^{(k)} - \xi_2^{(k)}\overline{\nabla}_{\B} (   \cL_{\bOmega, \tY}(\A^{(k)}, \B^{(k)}, \C^{(k)}) \nonumber \\
    &+ \lambda_2 r_2 (\bm B) ),\bm 0), \label{eq:btd_b}
\end{align}
where $\xi_1^{(k)}$ and $\xi_2^{(k)}$ are the step sizes for $\A$ and $\B$ updates respectively.
In practice, the step size parameters can be chosen using heuristics advocated in \texttt{Adagrad} or \texttt{Adam}, which are often effective. 
After the algorithm is terminated, the estimated radio map $\widehat{\tX}$ is obtained using
\[    \widehat{\tX} = \sum_{r=1}^R (\widehat{\A}_r \widehat{\B}_r^\top) \circ \widehat{\bc}_r, \]
where $\widehat{\A}_r, \widehat{\B}_r$ and $\widehat{\bm c}_r$ are the algorithm-found solutions.

Note that the procedure above is a typical inexact BCD approach, whose convergence properties were well studied and discussed in the literature; see \cite{hong2015unified}.
The algorithm described in \eqref{eq:btd_c}, \eqref{eq:btd_a} and \eqref{eq:btd_b} is referred to as {\it quantized spectrum cartography via BTD} (\texttt{QuantSC-BTD}).

\subsection{Algorithm For DGM-Based MLE}
Problem \eqref{eq:MLE} can be tackled using similar ideas. To see this, we consider the following reformulation:
\begin{align}\label{eq:MLE_nn}
    \minimize_{\C,\Z}&~ \cL_{\bOmega, \tY}(\Z, \C), \\ 
    {\rm subject~to}&~\bm C\geq \bm 0, \nonumber
\end{align}
where $$\cL_{\bOmega, \tY}(\Z, \C) =  F^{\rm DGM}_{\bm \varOmega,\tY}(\Z,\C) + \mu_1 q_1(\bm Z) + \mu_2 q_2(\bm C).$$
In practice, 
$q_1(\cdot):\mathbb{R}^{R \times D} \rightarrow \mathbb{R}$ is used to regularize $\Z$ (and thus also regularizing $\bm S_r = \bm g_{\theta}(\z_r)$). In this work, we set $$q_1(\Z)=\sum_{r=1}^R\|\bm z_r  \|_2^2=\|\bm Z\|_{\rm F}^2,$$ which would encourage the solution to reduce the Euclidean norm of the found SLFs, so that the specification in \eqref{eq:sln_set_CS} is more likely to hold.
Similar as before, we set $q_2(\C)=\|\C\|_{\rm F}^2$.
The nonnegativity constraint on $\bm C$ is again added to reflect its physical meaning. Note that $\bm Z$ needs not to be nonnegative and thus there is no such constraint on the latent embeddings. The SLFs are nonnegative and bounded. Hence, $\bm g_{\bm \theta}(\cdot)$ uses the sigmoid activation functions at the output layer.  Using sigmoid makes $\|\widehat{\S}_r\|_{\infty}\leq 1$. This does not lose generality or hurt the recoverability. 
To see this, assume that $\widehat{\S}_r =\S_r$ and $\widehat{\bm c}_r=\bm c_r$ are estimated perfectly. One can always let $\widehat{\S}_r = \nicefrac{\S_r}{\|\S_r\|_{\infty}}$ (so that $\|\widehat{\S}_r \|_\infty = 1$) and $\widehat{\bc}_r=\|\S_r\|_\infty\bc_r$. Such scaling/counter-scaling does not change the outer product; i.e., $\widehat{\S}_r \circ \widehat{\bm c}_r= \S_r \circ \bm c_r$ still holds---which means $\widehat{\tX}=\tX$.

The BCD algorithm for tackling Problem~\eqref{eq:MLE_nn} is similar as that for handling \eqref{eq:MLE_btd}, except that only two blocks are updated.
To be specific,
in the $k$th iteration, we first update $\C$ using gradient projection:
\begin{align}\label{eq:dgm_c}
    &\C^{(k+1)} \gets  \\
    &\max( \C^{(k)} - \upsilon_2^{(k)}\overline{\nabla}_{\C} (  \cL_{\bOmega, \tY}(\Z^{(k)}, \C^{(k)})  + \mu_2 q_2 (\bm C) ),\bm 0). \nonumber
\end{align}
Then, the $\Z$-update is carried out as follows:
\begin{align}\label{eq:dgm_z}
\Z^{(k+1)} \gets \Z^{(k)} &- \upsilon_1^{(k)} \overline{\nabla}_{\Z} ( \cL_{\bOmega, \tY}(\Z^{(k)}, \C^{(k)} ) \nonumber \\
&+\mu_1 q_1(\Z));
\end{align}
In this case, as a DGM is involved, the gradient-related directions w.r.t. $\Z$ can be found by back propagation-based methods, also using numerical gradient-finding tools.
Again, when the algorithm stops, we obtain the estimated radio map $\widehat{\tX}$ via
\[  \widehat{\tX} = \sum_{r=1}^R g_{\btheta}(\widehat{\z}_r) \circ \widehat{\bc}_r,  \]
where $\widehat{\z}_r$ and $\widehat{\bc}_r$ are the solutions found by the algorithm. 

The algorithm in \eqref{eq:dgm_c}-\eqref{eq:dgm_z} is referred to as the {\it quantized SC via DGM} (\texttt{QuantSC-DGM)}).

\begin{figure}
\includegraphics[width=0.9\linewidth]{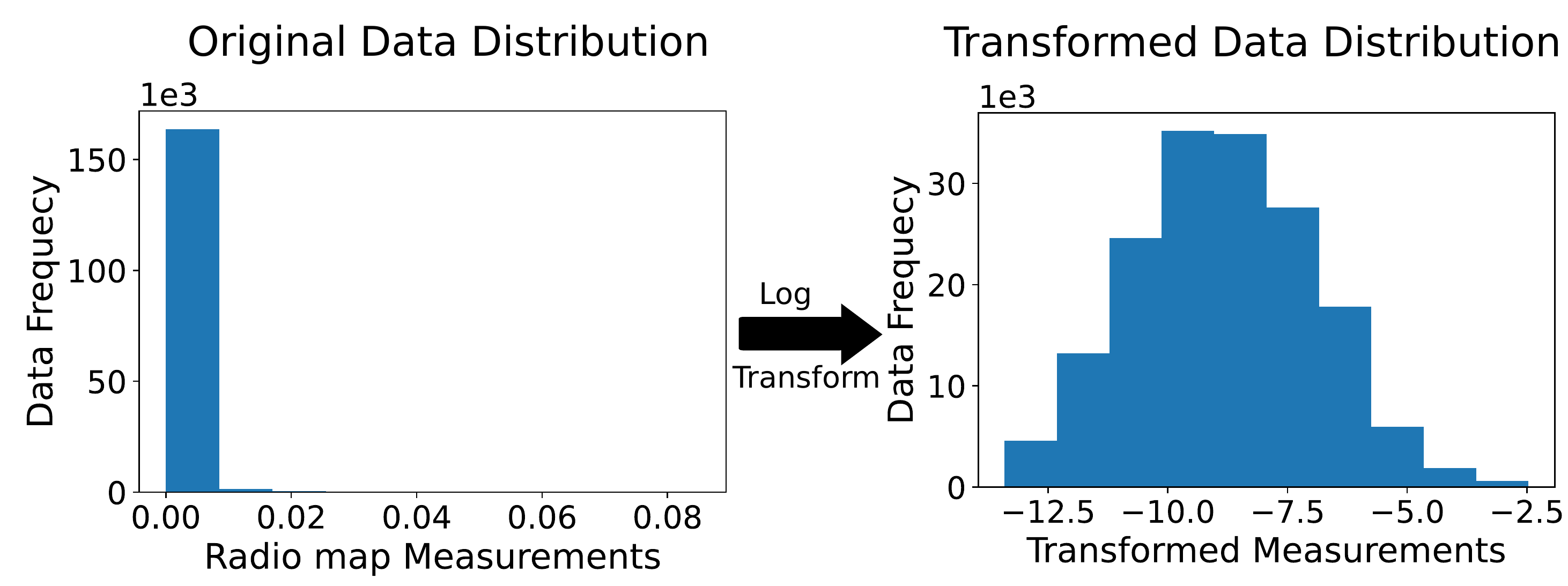}
\caption{Histograms of measurement values before (left) and after (right) the log-transform in \eqref{eq:logtrans}. The radio map covers a $51\times 51$m$^2$ region with $R=4$ emitters; $K=64$ frequencies; shadowing parameters are $(X_c,\eta)=(50,6)$ (see definitions in Sec.~\ref{section:synth_experiments}). }
\label{fig:data_visualization}
\end{figure}

\subsection{Quantization Interval Design}\label{section:quantization_design}
One challenge for quantizing radio maps is that the PSD values across space are extremely skewed; see Fig. \ref{fig:data_visualization}. Directly applying simple quantizers (e.g., the uniform quantizer) does not work well.
As mentioned, we propose to apply the $\log$-transformation in \eqref{eq:logtrans} before quantization.
Recall that the transformation is expressed as $h(x)=\log(x+a)$ with $a>0$, which ``condenses'' the dynamic range of the radio map. Log-transformation is commonly used in quantization and data pre-processing for nonnegative data; see, e.g., \cite{papalexakis2012k}.

After the log-transformation, the $b_q$'s in \eqref{eq:model_qsc} can be determined by making every interval roughly contain the same number of data points (i.e., $\tX(i,j,k)$). This strategy was employed in the matrix and tensor completion literature \cite{lan2014matrix, lan2013tag} and was proven effective. The intuition is that we want to make ``maximum utilization'' of all quantization levels.
In an SC system, the $\{b_q\}$ should be determined by the system designers prior to deployment.
Our suggested steps are as follows: 1) The system generates a set of simulated radio maps from a range of environment parameters that are considered reasonably reflecting the RF situation of the region of interest. 2) Using the simulated data, the system learns $b_q$ for $q\in [Q]$ using the following principle:
\begin{align*}
    b_q = \inf \left\{m \bigg | \frac{q-1}{Q-1} \leq F(m), m = h(x)  \right\}, \forall q = 2,\ldots,Q-1,
\end{align*} where $F(m)$ is the empirical CDF of the simulated $\tM$'s entries, and we set $b_0 = -\infty$ and $b_Q = +\infty$. We calculate bin boundaries from a large number (e.g., 1,000) of simulated radio maps and take average. 
 3) The fusion center broadcasts $\{b_1,\ldots,b_Q\}$ to all the sensors. Then, the sensors use the $b_q$'s to implement their quantizer.

\section{Recoverability Analysis}
In this section, we present the recoverability analyses of the formulated problems in \eqref{eq:btd_MLE} and \eqref{eq:MLE}, respectively.

\subsection{Technical Preparations}
It is readily seen that the following fact holds:
\begin{Fact}\label{fact:alpha}
    For $\tX\in {\cal X}^{\rm BTD}$ and $\tX\in {\cal X}^{\rm DGM}$, there always exist $\alpha^{\rm BTD} \in [0,\infty)$ and $\alpha^{\rm DGM}\in[0,\infty)$ such that
    \begin{align}
         \|\tX\|_{\infty} \leq \alpha,
    \end{align}                       
    where $\alpha = \alpha^{\rm BTD}$ if $\tX\in {\cal X}^{\rm BTD}$ and $\alpha=\alpha^{\rm DGM}$ if $\tX\in {\cal X}^{\rm DGM}$.
\end{Fact}
\begin{IEEEproof}
    The claim holds as the latent factors constituting $\tX$ in ${\cal X}^{\rm BTD}$ and ${\cal X}^{\rm DGM}$ are bounded; see \eqref{eq:sln_set_ABC} and \eqref{eq:sln_set_CS}.
\end{IEEEproof}

As our framework uses a log-transformation to pre-process the sensor-acquired data, we will need the following lemma in our later analysis:
\begin{Lemma}\label{lemma:trans_function}
Let $h(\tX)$ denote a tensor such that
$[h(\tX)]_{i,j,k} = \log(\tX(i,j,k)+a)$, where $\tX\geq \bm 0$ and $a>0$.
Suppose that $\| \tX \|_{\infty} \leq \alpha$. Then, for $\tX \neq \tX'$, the following holds:
\begin{align}\label{eq:lip}
    \frac{1}{\alpha + a} \leq \frac{\| h(\tX) - h(\tX') \|_{\rm F}}{\| \tX - \tX' \|_{\rm F}} \leq \frac{1}{a}.
\end{align}
\end{Lemma}
The lemma says that $h(\cdot)$ exhibits continuity when the argument is nonnegative.
The parameter $a>0$ is used to present the mapped value goes to $-\infty$, which can be an inconsequential small value (i.e., a value that does not dominate the input signal) in practice.
The proof is relegated to Appendix \ref{proof:lemma1}.
We will use the following fact:
\begin{Fact}\label{fact:constants}
Let ${\cal X}\in \{ {\cal X}^{\rm BTD}, {\cal X}^{\rm DGM}\} $ and ${\cal M} =\{ h(\tX)~|~\forall \tX\in {\cal X} \}.$
Then, if $b_\ell \neq b_j$ for any $\ell\neq j$ $\forall \ell, j \in [Q]$, there always exist bounded constants $\ualpha$,  $\lalpha$ and $\falpha$ such that
\begin{subequations}\label{eq:constants}
    \begin{align}
        \label{eq:u_alpha}
         \ualpha := & \sup_{m} \max_{\ell \in [Q]} \log \left( \frac{1}{ f_\ell(m) } \right ) \\
        \label{eq:l_alpha}
        \lalpha := & \sup_{m} \max_{\ell \in [Q]} \frac{\left| \dot f_\ell(m)\right|}{ f_\ell(m) }\\
        \label{eq:beta_alpha}
        \falpha := & \inf_{m} \max_{\ell \in [Q]} \frac{(\dot f_\ell(m))^2}{ f_\ell(m) },
\end{align}
\end{subequations}
 where $\alpha$ is as defined in Fact~\ref{fact:alpha}.
\end{Fact}

The proof of the Fact \ref{fact:constants} is presented in Appendix \ref{proof:fact1}.
Fact~\ref{fact:constants} means that the if our radio map measurements are bounded, then $f_\ell(\cdot)$ does not change too sharply or (for at least one $\ell\in [Q]$) too slowly, reflected in $\lalpha$ and $\falpha$, respectively.
The existence of these parameters makes the function locally analogous to functions that have gradient-Lipschitz continuity and strong convexity; see discussions in \cite{davenport20141, cai2013max}.

\subsection{Recoverability Under The BTD Model}
We first analyze the recoverability of \eqref{eq:MLE_btd}.
Before we proceed,
it should be mentioned that in the previous sections, 
for the simplicity of presentation,
we slightly abused the notations by using $\tX$, $\S_r$, $\bm c_r$, etc. to represent both the ground-truth parameters and the optimization variables (see, e.g., \eqref{eq:MLE_btd}).
In the sequel, for the clarity of analysis, we will use $\natural$ to denote the ground-truth parameters (e.g., $\C^\natural$), to distinguish them from the optimization variables (e.g., $\C$ in \eqref{eq:MLE_btd}).
Let us make the following assumption:
\begin{Assumption}\label{ass:approx_btd}
There exists a constant $\nu^{\rm BTD} \in [0,+\infty)$ such that 
\begin{align}
 \min_{{\tX}\in \cX^{\rm BTD}}\|\tX - \gtX \|_{\infty} \leq \nu^{\rm BTD},  
\end{align}
where $\tX^\natural$ represents the ground-truth radio map.
\end{Assumption}
The existence of $\nu^{\rm BTD}>0$ makes sense, which could be a result of sensing noise or just due to modeling errors.

\begin{Theorem} \label{thm:btd}
{Suppose that Assumption \ref{ass:approx_btd} holds. 
Assume that $\bOmega$ is uniformly sampled from $[I]\times [J]$ with replacement and that $N=|\bm \varOmega|$.
Define
$\tX^\star = \sum_{r=1}^R(\A^\star_r{\B^\star_r}^\top)\circ \bm c^\star_r$, where
\begin{align}
    (\A^\star, \B^\star, \C^\star) = \arg\min_{(\A,\B,\C) \in  {\cal S}^{\rm BTD}}~F^{\rm BTD}_{\bm \varOmega,\tY}(\A,\B,\C),
\end{align}
is an optimal solution of the {\sf MLE via BTD} criterion in \eqref{eq:btd_MLE}.
{Then, with probability at least $1-2\delta$, we have
    \begin{align} \label{eq:main_result_btd}
    \frac{\| \tX^\star - \gtX \|_{\rm F}^2}{IJK} \leq&  \frac{8 C_1 C_2 (1+\tau)}{K}\sqrt{\frac{R}{N}} + \ualpha C_1 \sqrt{\frac{\log(\frac{1}{\delta})}{2N}} \nonumber \\
    &  +  \ualpha C_1 \sqrt{ \frac{8\log(\frac{2}{\delta})}{N} } +  C_1 C_2\nu^{\rm BTD},
\end{align}}
    $ \text{where, } \tau = 3 \bigg( \sqrt{((I+J)L+K) \log ( 3\sqrt{R}(\beta + \kappa) )  }
    + \sqrt{ K\log(\kappa/2)+ {(I+J)L}\log(\beta)} \bigg ), $} and $\beta, \kappa$ are defined in \eqref{eq:sln_set_ABC}. The constants $C_1$ and $C_2$ are given by $C_1 = \nicefrac{4 (\alpha + a)^2}{\falpha},~{\rm and} ~ C_2 = \nicefrac{ \lalpha}{a}.$

\end{Theorem}
The proof is relegated to Appendix~\ref{Appendix:thm_proof}.
Theorem~\ref{thm:btd} shows that the {\it mean squared error} (MSE) of the {\it optimal} solution given by \eqref{eq:btd_MLE} is able to approach the ground-truth $\tX^\natural$, if $N=|\bm \varOmega|$ is sufficiently large and the tensor-based representation error (i.e., $\nu^{\rm BTD}$) is reasonably small.
Our result also shows that $N=\Omega(\nicefrac{(I+J)L+K)R \log ( 3\sqrt{R}(\beta + \kappa) )}{K^2} )$   is the sample complexity that suffices to ensure provable recovery, which is at the same order of the number of unknown parameters. This is consistent with the previous results from classic matrix/tensor completion literature under other low-rank models \cite{davenport20141, bhaskar2016probabilistic, cao2015categorical, ghadermarzy2018learning, li2018tensor}.

\begin{Remark}
The recoverability of low-rank tensors under 1-bit quantization was studied in \cite{ghadermarzy2018learning}, but was under the canonical polyadic decomposition (CPD) model.
Multi-level quantization based tensor completion was considered in 
\cite{ghadermarzy2018learning, li2018tensor, lee2020tensor}, but again under tensor models such as the CPD and Tucker models. In this work, we formulate our problem under the BTD model, which is more suitable for the SC problem. 
Recoverability of radio map tensors under the BTD model was considered in \cite{zhang2020spectrum} under a different setting. There, the sampling pattern was deterministic and no quantization was involved (a similar setting was used in the context of hyperspectral super-resolution \cite{ding2020hyperspectral}). 
Hence, their proof took a very different route via using the essential uniqueness of BTD, which is not needed in our case.

 The random sampling and quantization settings in \cite{ghadermarzy2018learning, li2018tensor, lee2020tensor} are closer to ours, but the data acquisition in these works was done through independently sampling the entries of the tensor. However, in this work, we focus on the scenario where the sensors can sample the entire PSD $\tX(i,j,:)$, also known as the {\it fiber sampling} case \cite{fu2020block}. 
Our proof extends those in \cite{ghadermarzy2018learning, li2018tensor, lee2020tensor} and carefully accommodates these differences.
\end{Remark}

\subsection{Recoverability Under The DGM Model}
The proof under the DGM of the radio map \eqref{eq:MLE} follows a similar technical route, except that the DGM's ``complexity measure'' needs to be carefully derived. As in the previous section, we make the following assumption:

\begin{Assumption}\label{ass:approx_dgm} There exists a constant $\nu^{\rm DGM} \in[0,+\infty)$ such that
\begin{align}
   \min_{{\tX}\in \cX^{\rm DGM}}\|{\tX} - \gtX \|_{\infty} \leq \nu^{\rm DGM}.
\end{align}
\end{Assumption}
Ideally, if $\bm g_{\bm \theta}(\cdot)$ is a universal function approximator, $\nu^{\rm DGM}$ would have been zero (provided that other types of noise do not exist).
However, in practice, $\bm g_{\bm \theta}(\cdot)$ is represented by a neural network with finite depth/width, which means that it is never universal and thus a nonzero $\nu^{\rm DGM}$ exists.
Note that $\nu^{\rm DGM}$ can be decreased if the neural network $\bm g_{\bm \theta}(\cdot)$ is more expressive. A more expressive neural network often admits a deeper and wider neural architecture, which has a increased number of parameters.

We also make the following structural assumption:
\begin{Assumption}\label{ass:generator_structure}
The generator network is an $L$-layer neural network 
\begin{align}\label{eq:deepnet}
\g_{\btheta}(\z) = \text{mat}(\bm \zeta_L(\A_L(...\bm \zeta_1(\A_1\z)))),
\end{align}
 where ${\rm mat}(\cdot):\mathbb{R}^{IJ}\rightarrow \mathbb{R}^{I\times J}  $
reshapes a vector to a matrix (i.e., the inverse operation of vectorization), $\A_\ell \in \bbR^{D_\ell \times D_{\ell-1}}$ is the network weight in the $\ell$th layer, in which $D_0 = D$ and $D_L = IJ$, $\bm \zeta_\ell(.) = [\zeta(.),...,\zeta(.)]^T : \bbR^{D_\ell} \to \bbR^{D_\ell}$ is a $\phi_\ell$-Lipschitz activation function with $\zeta(.): \bbR \to \bbR$, and $P= \prod_{\ell=1}^L \phi_\ell \| \A_\ell \|_2 < \infty.$
In addition, the latent embedding $\z$ is from a bounded set, i.e., $\z\in \mathcal{Z} = \{\z \in \bbR^D : \| \z \|_2 \leq q\}$. 
\end{Assumption}
Assumption~\ref{ass:generator_structure} follows the same setting in \cite{shrestha2022deep}.
This assumption characterizes the structure of generative neural network. Note that many commonly used activation functions are Lipschitz continuous, e.g., ReLU, sigmoid, and tanh. In addition, \eqref{eq:deepnet} subsumes many deep network architectures as its special cases, e.g., the multi-layer perceptron (MLP) and the convolutional neural network (CNN).

With the assumptions, we show the following recoverability theorem:
\begin{Theorem} \label{thm:dgm}
Suppose that Assumption~\ref{ass:approx_dgm} and \ref{ass:generator_structure} hold.
Assume that $\bOmega$ is uniformly sampled from $[I] \times [J] $ and that $|{\bm \varOmega}| = N$.
Define $\tX^\star = \sum_{r=1}^R \bm g_{\bm\theta}(\z^\star_r)\circ \bm c_r^\star$, where
\begin{align}
    (\C^\star,\Z^\star) = \arg\min_{(\C,\Z)\in \cS^{\rm DGM} } ~F^{\rm DGM}_{\bm \varOmega,\tY}(\C,\Z),
\end{align} is an optimal solution of the {\sf MLE via DGM} criterion in \eqref{eq:MLE}.
Then with probability at least $1-2\delta$, we have
    \begin{align}\label{eq:main_result_dgm}
    \frac{\| \tX^\star - \gtX \|_{\rm F}^2}{IJK} \leq&  \frac{8 C_1 C_2(1+\tau)}{K}\sqrt{\frac{R}{N}} + \ualpha C_1 \sqrt{\frac{\log(\frac{1}{\delta})}{2N}} \nonumber \\
    &  +  \ualpha C_1 \sqrt{ \frac{8\log(\frac{2}{\delta})}{N} } + C_1 C_2 \nu^{\rm DGM},
\end{align}
    where \begin{align*}
    \tau = 3\sqrt{K \log ( \frac{3}{2}\sqrt{R}\kappa(\beta + \kappa) )+ D\log ( 3\sqrt{R}Pq(\beta + \kappa) ) },
\end{align*}
 and $\beta, \kappa$ are defined in \eqref{eq:sln_set_CS}. The constants $C_1$ and $C_2$ are given by $C_1 = \nicefrac{4 (\alpha + a)^2}{\falpha},~{\rm and} ~ C_2 = \nicefrac{\lalpha}{a}.$
\end{Theorem}
The proof of Theorem \ref{thm:dgm} is presented in Appendix \ref{Appendix:thm_proof}. Theorem \ref{thm:dgm} shows that the optimization criterion in \eqref{eq:MLE} ensures the recoverability of the ground-truth radio map $\gtX$---even when the radio map is parameterized by a learned DGM other than a classic low-rank model.
There are two primary sources of errors in our recovery process. The first error occurs due to the limited fiber samples of radio maps measurement, also referred to as generalization (or statistical) error, which is given by the first term on the R.H.S. of \eqref{eq:main_result_dgm}. The second type of error is induced by the representation error of the DGM, i.e., $\nu^{\rm DGM}$, which is the last term in \eqref{eq:main_result_dgm}.
One can see that the recovered radio map becomes closer to the ground truth with an increased sample size at the rate of $\cO(1/\sqrt{N})$. When the DGM becomes more complex (e.g., when the depth/width of network is increased), the approximation error $\nu^{\rm DGM}$ decreases.  However, the increased network complexity makes $\tau$ larger. This presents a reasonable trade-off between model and sample complexities.
In short, one hopes to use a reasonably complex DGM so that $\nu^{\rm DGM}$ is small, but not an overly expressive DGM that might lead to over-fitting effects.

\section{Experiments}

\subsection{Synthetic Data Experiments}\label{section:synth_experiments}

\subsubsection{Data Generation}
For data generation, we first consider the 2D geographical region of $50 \times 50~{\rm m}^2$ that has $51 \times 51$ grids. The region is discretized using indices $\{0, 1, \ldots 50\} \times \{0, 1, \ldots 50\}$ such that $I= J= 51$. The spectral domain has $K=64$ subbands. We generate the radio maps by first generating the SLF of each emitters using the joint path loss model and spatial correlated log-normal shadowing model \cite{goldsmith2005wireless}. 
The two key parameters for this model are the decorrelation distance $X_c$ and the shadowing variance $\eta$. Larger $X_c$ and smaller $\eta$ correspond to milder shadowing environment; see \cite{shrestha2022deep} for more details. The PSDs are also generated following the same method as in \cite{zhang2020spectrum,shrestha2022deep}.

\subsubsection{DGM Learning}\label{sec:dgmlearning}
We learn the DGM using GAN. We set $D = 256$. The latent embedding $\z_r$ is sampled from the standard Gaussian distribution. The detailed architecture of the GAN is in Appendix \ref{Appendix:architecture}.
To generate training samples for learning the DGM, we create 5,000 random SLFs using random emitter locations, a range of path loss coefficients, $X_c$ and $\eta$ in [2,2.5], [30,100] and [3,8], respectively.
We then train our GAN using \texttt{Adam} optimizer with a batch size of 128 with 250 epochs. The initial learning rates for the discriminator and generator are set to be 0.0004 and 0.0001 respectively.

\subsubsection{Algorithm Settings}
We initialize $\A$ and $\B$ as zero matrices in the BTD model. For the DGM model, we initialize $\z_r$ using the standard normal distribution. For both models, we initialize the elements of $\C$ using the uniform distribution between 0 and 1. 
Under both the DGM and BTD models, we stop the algorithms when the relative change of the cost function is smaller than $10^{-3}$ or when the BCD algorithms reach 300 iterations.
For both of our reformulations in \eqref{eq:MLE_nn} and \eqref{eq:MLE_btd}, the regularization terms are all set to be $10^{-3}$. We set $L=10$ for the BTD model following \cite{zhang2020spectrum}.
For optimization, we use the \texttt{Adam} optimizer for both DGM and BTD block updates. For both models, we set the initial learning rates for $\bm C$ to be $ 0.003$ and for the other variables to be $0.006$.

\subsubsection{Quantization} We set $a = 10^{-6}$ in $h(x) = \log(x+a)$. For quantization, we set the noise variance using a validation set consisting of 20 simulated radio maps. The validation set is generated with shadowing parameters $X_c\in [30, 100]$ and $\eta\in [3.0, 8.0]$. The recovering algorithms are applied to the validation set with various $\sigma^2\in [0.5, 4.0]$ for different bits cases and the average best-performing $\sigma^2=1.7$ is selected.
The quantization bins are constructed using $1,000$ simulated radio maps data generated following the description in Sec.~\ref{sec:dgmlearning}. The $b_q$'s are constructed using the method in Sec.~\ref{section:quantization_design} for $B \in [1, 8]$ , where $B= \log_2(Q-1)$ bits.

\subsubsection{Performance Metrics}
We use \textit{log-domain normalized reconstruction error} (LNRE) as our metric: 
\begin{align}
    {\rm LNRE} = \frac{\| \widehat{\tM} - \gtM \|^2_{\rm F}}{\| \gtM \|^2_{\rm F}},
\end{align}
where $\widehat{\tM} = h(\widehat{\tX})$ is the estimated log-transformed the radio map. The LNRE is an appropriate metric for skewed data as it prevents a few larger values in the $\tX$ from dominating the performance evaluation. All the LNREs are averaged over 10 random trials.

\subsubsection{Baselines} We use the DGM-based SC method in \cite{shrestha2022deep}, namely, ``\texttt{DowJons}'', as it demonstrated state-of-the-art SC performance given that real-valued feedback can be used.
We also use the BTD-based method for unquantized data in \cite{zhang2020spectrum}. Additionally, we use the method in \cite{romero2017learning} based on kernel regression for quantized SC (referred to as ``\texttt{KR}''). Note that \cite{romero2017learning} assumes that the PSDs of the emitters are known to the fusion center, but our method does not need this information.

\subsubsection{Results}
\begin{figure}[t!]
    \centering
    \includegraphics[width=1\linewidth]{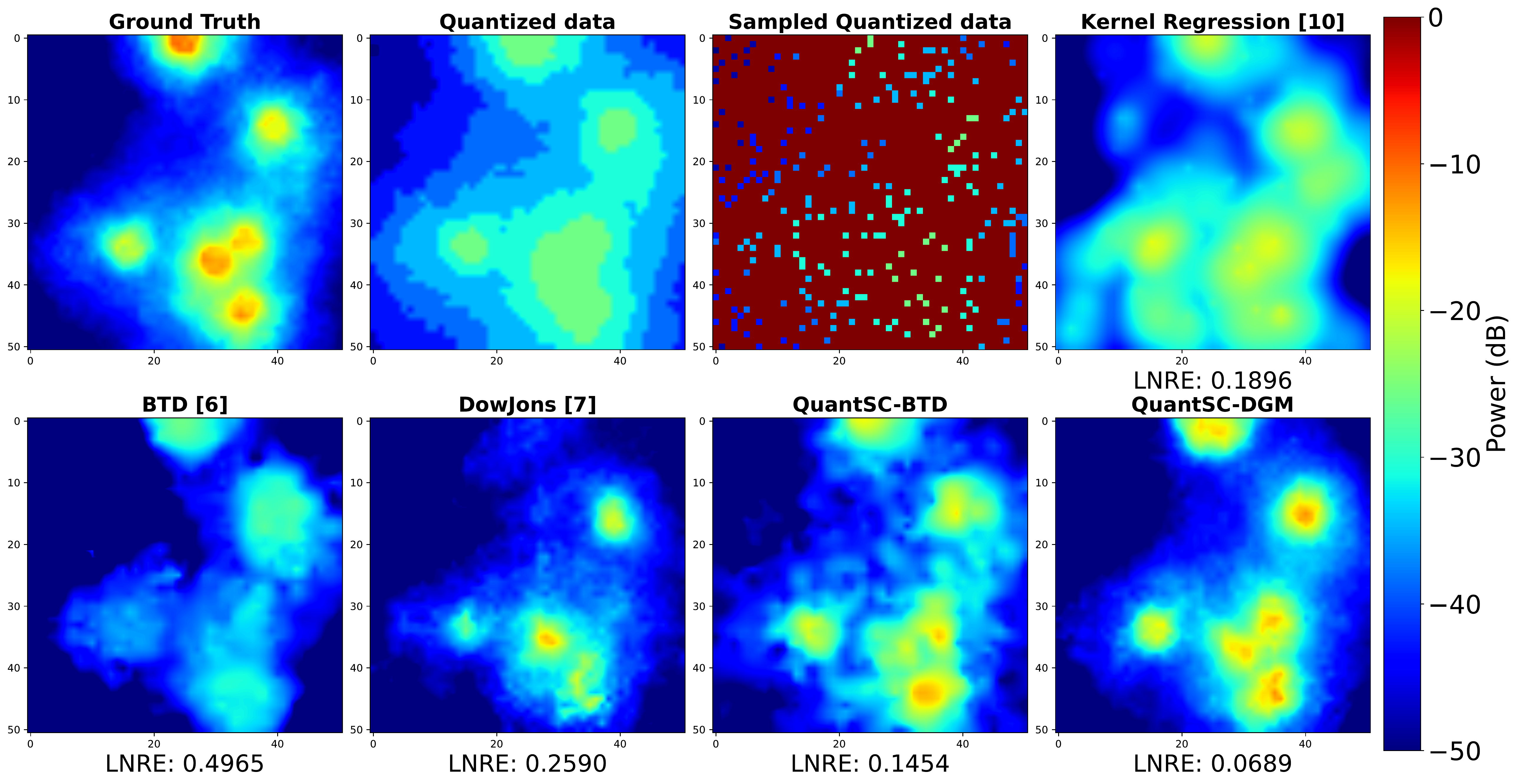}    \caption{ Ground-truth and reconstructed radio maps by various methods at the 30th frequency bin; $\rho = 10
    \%$, $R=6$, $X_c =50$, $\eta =6$ and $\sigma^2 =1.7$, $B=3$ bits. } 
    \label{fig:comparison_no_shadow}
\end{figure}

Fig. \ref{fig:comparison_no_shadow} shows an illustrative example where we have $R=6$ emitters. For visualization, we show the recovery result at a single frequency bin.
In this experiment, $\rho = 10 \%$ of the $51 \times 51$ grids are sampled, where $\rho=|\bm \varOmega|/IJ\times 100\%$. We use $B=3$ bits for quantizing every real-valued measurement. The shadowing parameters are set to be $X_c =50$ and $\eta =7$. The LNRE of the reconstructed radio maps over all frequencies are shown along with each images. One can see that the \texttt{QuantSC-DGM} offers the best recovery performance, both visually and in terms of LNRE.
The \texttt{QuantSC-BTD} works reasonably, as it still recovers the positions of emitters well. The \texttt{DowJons} method clearly missed some emitters, which is not surprising---as it is not designed for quantized SC.The BTD-based method from \cite{zhang2020spectrum} also performs poorly as it also does not consider quantized measurements. The \texttt{KR} method did not perform well and over-smoothing effects are observed.

\begin{figure}[t!]
    \centering
    \includegraphics[width=1\linewidth]{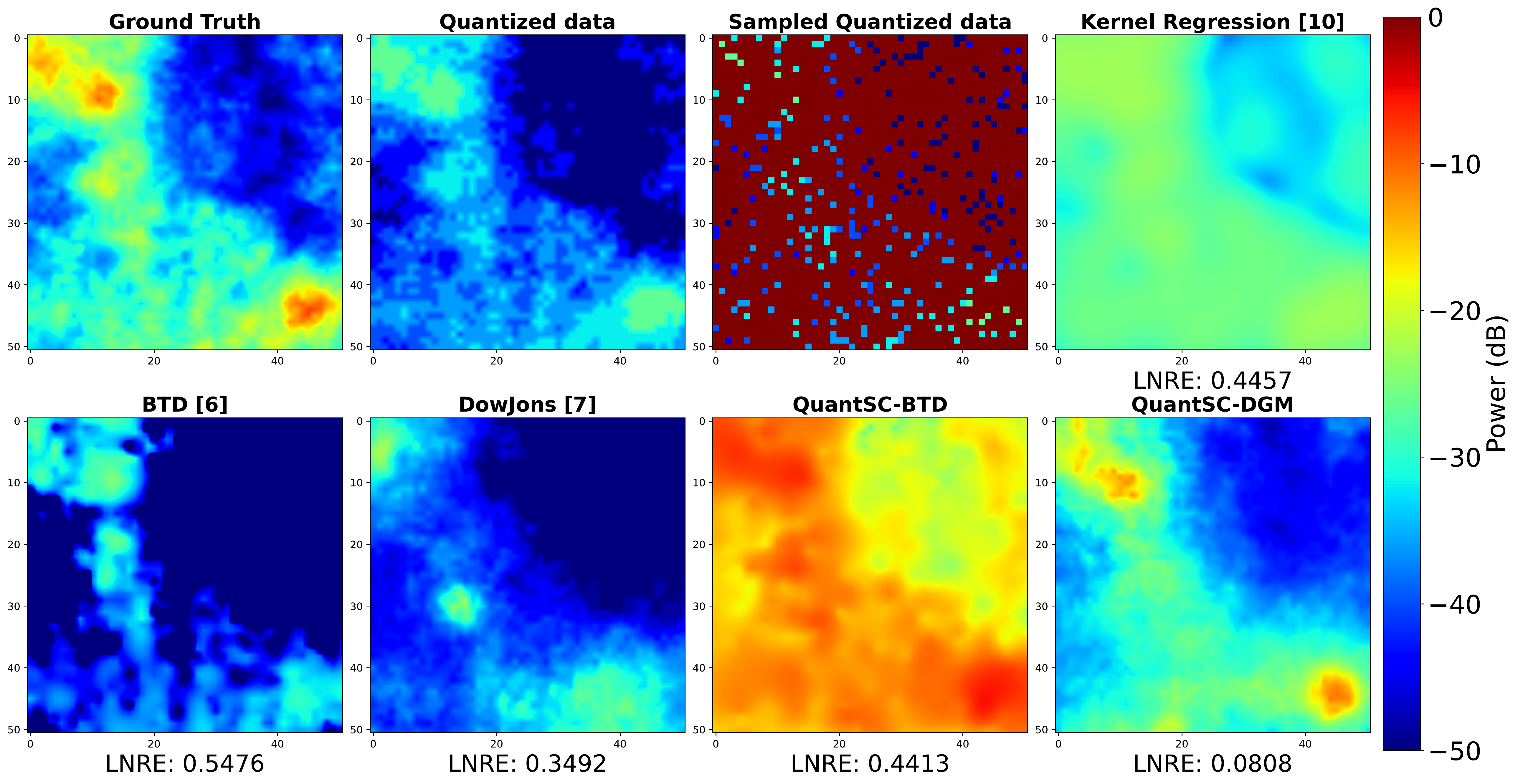}    \caption{ Ground-truth and reconstructed radio maps under heavy shadowing by various methods at the 20th frequency bin; $\rho = 10
    \%$, $R=2$, $X_c =40$, $\eta =8$ and $\sigma^2 =1.7$, $B=3$ bits.}
    \label{fig:comparison_shadow}
\end{figure}

Fig. \ref{fig:comparison_shadow} shows the result with heavier shadowing compared to the scenario in Fig.~\ref{fig:comparison_no_shadow}.
Here, we use $R=2$ and set $X_c$ and $\eta$ to be $40$ and $8$, respectively. All the other settings follow those in in previous paragraph. 
Clearly, this is a much more challenging scenario as most of the methods fail.
\texttt{QuantSC-BTD} does not work well in this case, as expected---since the low-rank assumption of the SLFs is grossly violated.
The proposed \texttt{QuantSC-DGM} is the only one that produces a radio map close to the ground truth. This shows the power of DGM in terms of modeling complex environments.

\begin{table}[t!]
\begin{center}
\caption{ LNRE under various $\eta$'s: $R=6$, $X_c = 50$, $\rho= 10\%$, and $B=3$ bits. }
\label{table:eta}
\resizebox{\linewidth}{!}{
\begin{tabular}{| c | c | c | c | c |}
\hline
 $\eta$ & \texttt{KR}\cite{romero2017learning} & \texttt{DowJons}\cite{shrestha2022deep} & \texttt{QuantSC-BTD} & \texttt{QuantSC-DGM}\\
\hline
4.0 & 0.1351 \tpm 0.0092  & 0.2078 \tpm 0.0149 & 0.1352 \tpm 0.0082 & \textbf{ 0.0552 \tpm 0.0083} \\
\hline
5.0 & 0.1284 \tpm 0.0153 & 0.2163 \tpm 0.0058 & 0.1384 \tpm 0.0142 & \textbf{ 0.0650 \tpm 0.0041 } \\
\hline
6.0 & 0.1949 \tpm 0.0102 & 0.2435 \tpm 0.0163 & 0.1327 \tpm 0.0092 & \textbf{ 0.0665 \tpm 0.0103 } \\
\hline
7.0 & 0.2310 \tpm 0.0210 & 0.2341 \tpm 0.0132 & 0.1461 \tpm 0.0097 & \textbf{ 0.0701 \tpm 0.0084 } \\
\hline
8.0 & 0.2705 \tpm 0.0076 & 0.2578 \tpm 0.0097 & 0.1593 \tpm 0.0085 & \textbf{ 0.0787 \tpm 0.0069} \\
\hline
\end{tabular}}
\end{center}
\end{table}

\begin{table}[t!]
\begin{center}
\caption{LNRE under various $X_c$'s: $R=6$, $\eta = 6$ $\rho= 10\%$, and $B=3$ bits.}
\label{table:decorr}
\resizebox{\linewidth}{!}{
\begin{tabular}{| c | c | c | c | c |}
\hline
 $X_c$ & \texttt{KR}\cite{romero2017learning} & \texttt{DowJons}\cite{shrestha2022deep} & \texttt{QuantSC-BTD} & \texttt{QuantSC-DGM}\\
\hline
30.0 & 0.2366 \tpm 0.0303 & 0.2116 \tpm 0.0142 & 0.1513 \tpm 0.0093 & \textbf{ 0.0699 \tpm 0.0068 } \\
\hline
50.0 & 0.1852 \tpm 0.0095 & 0.2192 \tpm  0.0102 & 0.1316 \tpm 0.0129 & \textbf{ 0.0625 \tpm 0.0041 } \\
\hline
70.0 & 0.1567 \tpm 0.0162  & 0.2218 \tpm 0.0092 & 0.1318 \tpm 0.0087 & \textbf{ 0.0620 \tpm 0.0080 } \\
\hline
90.0 & 0.1401 \tpm 0.0095 & 0.1783 \tpm 0.0105  & 0.1185 \tpm 0.0123 & \textbf{ 0.0579 \tpm 0.0091 } \\
\hline
\end{tabular}}
\end{center}
\end{table}

Tables \ref{table:eta} and \ref{table:decorr} show the performance of the methods under various $\eta$ and $X_c$, respectively.
Recall that the two parameters reflect the level of shadowing. 
In all cases, \texttt{QuantSC-DGM} exhibits the lowest LNRE. The performance only deteriorates gracefully when $X_c$ decreases and $\eta$ increases, showing the method's robustness to heavy shadowing.
The \texttt{QuantSC-BTD} approach offers the second-best LNRE performance in all cases. As expected, it works the best when $X_c=90$ and $\eta=6$, but suffers from a 40\% LNRE increase when $X_c$ reduces to 30. This suggests that the shadowing heavily affects the performance of the BTD-based approach, as mentioned in \cite{shrestha2022deep} as well.

\begin{table}[!t]
\begin{center}
\caption{LNRE under various $R$'s. $\eta=6$, $X_c = 50$,  $\rho= 10\%$, and $B=3$ bits.}
\label{table:R}
\resizebox{\linewidth}{!}{
\begin{tabular}{| c | c | c | c | c |}
\hline
 $R$ & \texttt{KR}\cite{romero2017learning} & \texttt{DowJons}\cite{shrestha2022deep} & \texttt{QuantSC-BTD} & \texttt{QuantSC-DGM}\\
\hline
2 & 0.0818 \tpm 0.0193 & 0.1555 \tpm 0.0144 & 0.1125 \tpm 0.0241 & \textbf {  0.0499 \tpm 0.0036  } \\
\hline
4 & 0.1300 \tpm 0.0324 & 0.2515 \tpm 0.0404 & 0.1301 \tpm 0.0101 & \textbf {  0.0612 \tpm 0.0130  } \\
\hline
6 & 0.1730 \tpm 0.0078 & 0.2412 \tpm 0.0072 & 0.1442 \tpm 0.0129 & \textbf {  0.0640 \tpm 0.0087  } \\
\hline
8 & 0.1902 \tpm 0.0309 & 0.2746 \tpm 0.0315 & 0.1621 \tpm 0.0137 & \textbf {  0.0595 \tpm 0.0054  } \\
\hline
10 & 0.2295 \tpm 0.0639 & 0.3227 \tpm 0.0206 & 0.1860 \tpm 0.0160 & \textbf {  0.0659 \tpm 0.0071  } \\
\hline
12 & 0.2197 \tpm 0.0127 & 0.3522 \tpm 0.0433 & 0.1872 \tpm 0.0111 & \textbf {  0.0674 \tpm 0.0069  } \\
\hline
14 & 0.2181 \tpm 0.0099 & 0.3280 \tpm 0.0052 & 0.1985 \tpm 0.0181 & \textbf {  0.0746 \tpm 0.0061  } \\
\hline
16 & 0.2149 \tpm 0.0083 & 0.3321 \tpm 0.0056 & 0.1951 \tpm 0.0031 & \textbf {  0.0782 \tpm 0.0062  } \\
\hline
18 & 0.2296 \tpm 0.0167 & 0.3632 \tpm 0.0070 & 0.2148 \tpm 0.0076 & \textbf {  0.0782 \tpm 0.0041  } \\
\hline
20 & 0.2356 \tpm 0.0174 & 0.3630 \tpm 0.0388 & 0.2117 \tpm 0.0075 & \textbf {  0.0834 \tpm 0.0064  } \\
\hline
\end{tabular}}
\end{center}
\end{table}

Table \ref{table:R} shows the LNREs of the reconstructed radio maps under various $R$'s, i.e., the number of emitters. All the methods see performance degradation when $R$ increases, as more unknown parameters need to be estimated. However, the \texttt{QuantSC-DGM} and \texttt{QuantSC-BTD} still offer the best and second-best LNRE performance.

\begin{table}[!t]
\begin{center}
\caption{ LNRE under various $\widehat{R}$'s. Ground-truth $R=6$, $X_c = 40$, $\eta = 7.0$, $\rho= 10\%$, and $B=3$ bits. }
\label{table:estimate_R}
\resizebox{\linewidth}{!}{
\begin{tabular}{| c | c | c | c | c |}
\hline
 $\widehat{R}$ & \texttt{KR}\cite{romero2017learning} & \texttt{DowJons}\cite{shrestha2022deep} & \texttt{QuantSC-BTD} & \texttt{QuantSC-DGM}\\
\hline
2 & - & 0.3058 \tpm 0.0029 & 0.1790 \tpm 0.0098 & \textbf{ 0.0963 \tpm 0.0077 }  \\
\hline
3 & - & 0.2931 \tpm 0.0047 & 0.1731 \tpm 0.0085 & \textbf{ 0.0852 \tpm 0.0068 }  \\
\hline
4 & - & 0.2837 \tpm 0.0107 & 0.1706 \tpm 0.0097  & \textbf{ 0.0811 \tpm 0.0076 }  \\
\hline
5 & - & 0.2447 \tpm 0.0617 & 0.1639 \tpm 0.0101 & \textbf{ 0.0695 \tpm 0.0049 }  \\
\hline
6 & 0.1917 \tpm 0.0068 & 0.2398 \tpm 0.0090 & 0.1602 \tpm 0.0091 & \textbf{ 0.0682  \tpm 0.0083 }  \\
\hline
7 & - & 0.2415 \tpm 0.0115 & 0.1612 \tpm 0.0105& \textbf{ 0.0661 \tpm 0.0056 }  \\
\hline
8 & - & 0.2439 \tpm 0.0102 & 0.1610 \tpm 0.0102& \textbf{ 0.0673 \tpm 0.0083 }  \\
\hline
9 & - & 0.2956 \tpm 0.0114 & 0.1736 \tpm 0.0098 & \textbf{ 0.0696 \tpm 0.0137 }  \\
\hline
10 & - & 0.4474 \tpm 0.0220 & 0.1958 \tpm 0.0105 & \textbf{ 0.0716 \tpm 0.0093  } \\
\hline
11 & - & 0.4411 \tpm 0.0302 & 0.2017 \tpm 0.0107 & \textbf{ 0.0727 \tpm 0.0096 }  \\
\hline
\end{tabular}}
\end{center}
\end{table}

The simulation in Table. \ref{table:estimate_R} tests the robustness to the wrongly estimated $R$'s.
In the previous simulations, we assumed that the number of emitters $R$ is accurately estimated by the algorithms---which may not always hold in practice.
Table. \ref{table:estimate_R} shows the performance of all the methods when $R$ is underestimated or overestimated---i.e., the performance under wrong $R$'s. 
For clarity, we use $\widehat{R}$ to denote the ``estimated number of emitters'' that is used by the algorithms---and our purpose is to observe how the algorithms behave when $\widehat{R}\neq R$.
The true $R$ for this experiment setting is 6.
The \texttt{KR} method needs to know the PSD of the emitters, and thus we use the $\widehat{R}=R$ as its input.
One can see that when $\widehat{R} < R$, the algorithms have higher LNREs. The LNRE decreases when $\widehat{R}$ grows from 2 to 6, but increases again when $\widehat{R}\geq 9$. This makes sense as an underestimated $\widehat{R}$ could not capture all the ``principal components'' of the radio map tensor, but an overestimated $\widehat{R}$ makes the complexity of the model (represented by $\tau$ in Theorems~\ref{thm:btd} and \ref{thm:dgm}) higher. Hence, both seriously underestimated and largely overestimated $\widehat{R}$ could hurt the recovery accuracy. 
Similar phenomenon is also observed by changing the algorithm-used $\widehat{L}$ in the BTD (which may be different from the ground-truth $L$). The experiment using varying $\widehat{L}$ is not included due to page limitations, but the insights behind are the same.

\begin{table}[!t]
\begin{center}
\caption{LNRE under various $\rho$'s. $R=6$, $X_c = 50$, $\eta=6$, and $B=3$ bits. }
\label{table:rho}
\resizebox{\linewidth}{!}{
\begin{tabular}{| c | c | c | c | c |}
\hline
 $\rho$ & \texttt{KR}\cite{romero2017learning} & \texttt{DowJons}\cite{shrestha2022deep} & \texttt{QuantSC-BTD} & \texttt{QuantSC-DGM}\\
\hline
3\% & 0.2104 \tpm 0.0452 & 0.2625 \tpm 0.0163 & 0.1980 \tpm 0.0074 & \textbf{ 0.0900 \tpm  0.0070 } \\
\hline
5\% & 0.1975 \tpm 0.0201  & 0.2506 \tpm 0.0141   & 0.1756 \tpm 0.0106 & \textbf{ 0.0761 \tpm 0.0091} \\
\hline
10\% & 0.1722 \tpm 0.0124 & 0.2388 \tpm 0.0080 & 0.1429 \tpm 0.0116 & \textbf{ 0.0607 \tpm 0.0065 } \\
\hline
15\% & 0.1663 \tpm 0.0149 & 0.2220 \tpm 0.0079 & 0.1408 \tpm 0.0090 & \textbf{ 0.0557 \tpm 0.0083 } \\
\hline
20\% & 0.1416 \tpm 0.0075 & 0.1906 \tpm 0.0118  & 0.1315 \tpm 0.0071 & \textbf{ 0.0522 \tpm 0.0075 } \\
\hline
\end{tabular}}
\end{center}
\end{table}

Table \ref{table:rho} shows the performance of the algorithms under various $\rho$'s.
Notably, the LNRE output by \texttt{QuantSC-DGM} using $\rho=3\%$ is lower than the LNREs output by the other methods using $\rho=20\%$---again showing the expressive power of the DGM. However, the same DGM-empowered method \texttt{DowJons} performs worse compared to \texttt{QuantSC-BTD} under all $\rho$'s. This shows the effectiveness of our proposed quantized SC framework based on Gaussian quantization and MLE.

\begin{table}[!t]
\begin{center}
\caption{LNRE under various $B$'s. $R=6$, $X_c = 50$, $\eta=6$, and $\rho = 10\%$. }
\label{table:bits}
\resizebox{\linewidth}{!}{
\begin{tabular}{| c | c | c | c | c |}
\hline
 B & \texttt{KR}\cite{romero2017learning} & \texttt{DowJons}\cite{shrestha2022deep} & \texttt{QuantSC-BTD} & \texttt{QuantSC-DGM}\\
\hline
1 & 0.4670 \tpm 0.0059 & 0.7424 \tpm 0.0640 & 0.1815 \tpm 0.0112 & \textbf{ 0.0720 \tpm 0.0095 } \\
\hline
2 & 0.2885 \tpm 0.0260 & 0.5625 \tpm 0.0292 & 0.1548 \tpm 0.0159 & \textbf{ 0.0675 \tpm 0.0074 } \\
\hline
3 & 0.1781 \tpm 0.0082 & 0.2461 \tpm 0.0292 & 0.1434 \tpm 0.0140 & \textbf{ 0.0645 \tpm 0.0088 } \\
\hline
4 & 0.1547 \tpm 0.0114 & 0.1533 \tpm 0.0127 & 0.1385 \tpm 0.0132 & \textbf{ 0.0607 \tpm 0.0124 } \\
\hline
5 & 0.1384 \tpm 0.0057 & 0.0712 \tpm 0.0142 & 0.1339 \tpm 0.0168 & \textbf{ 0.0572 \tpm 0.0136 } \\
\hline
6 & 0.1128 \tpm 0.0100 & 0.0556 \tpm 0.0147 & 0.1242 \tpm 0.0078 & \textbf{ 0.0545 \tpm 0.0096 } \\
\hline
7 & 0.0970 \tpm 0.0079 & 0.0543 \tpm 0.0155 & 0.1028 \tpm 0.0119 & \textbf{ 0.0504 \tpm 0.0115 } \\
\hline
8 & 0.0743 \tpm 0.0090 & 0.0523 \tpm 0.0146 & 0.0972 \tpm 0.0109 & \textbf{ 0.0445 \tpm 0.0118 } \\
\hline
\end{tabular}}
\end{center}
\end{table}

Table \ref{table:bits} shows the performance under various numbers of quantization bits, i.e., $B$, used for each measurement. Compared to the baselines, the proposed methods \texttt{QuantSC-DGM} and \texttt{QuantSC-BTD} admit tangible margins when $B=1,2,3,4$. The performance of the proposed methods are more than satisfactory when even only $B=1$ bit is used. When $B\geq 5$, which means the quantization level reaches $Q=31$, the performance of \texttt{KR} and \texttt{DowJons} catch up---as the quantization error gradually becomes negligible in these cases.

\subsection{Real-Data Experiments}
\subsubsection{Data Description}
The data was obtained within a $14 \times 34 {\rm m}^2$ indoor space on an office floor at the Mannheim University. The data was acquired across 9 distinct frequency bands centered at 2.412 GHz, 2.422 GHz, 2.427 GHz, 2.432 GHz, 2.437 GHz, 2.442 GHz, 2.447 GHz, 2.457 GHz, and 2.462 GHz, respectively \cite{realdataset}. The region is divided into $1 \times 1 {\rm m}^2$ grids, and 166 of these grids were installed with sensors at their centers which are placed throughout the hallway.  
The leftmost column in Fig.~\ref{fig:real_data} displays the ground-truth radio map over the nine frequencies.
The white blocks in the figures are rooms where measurements could not be taken. 
The measurements in our experiments are uniformly sampled from the hallway area.
This type of mixed deterministic (rooms) and random (hallway) missing pattern is not exactly covered by the conditions made in our recoverability theorems, but the dataset can still be employed to test the algorithms. The results can indicate their usefulness in real-world settings and robustness to violation of the conditions in the theorems.

\subsubsection{Hyperparameter Settings}
The data is heavily skewed for the real data and the values are extremely small. The largest power measured is -40 dB. Hence, we take $a=10^{-15}$ for our transform to avoid $a$ dominating the output of $h(x+a)$.
For quantization, we follow the same validation-based strategy as before:
We set $\sigma^2 =4.0$, which is tuned using a validation set consisting of 20 simulated radio maps. The validation set is generated with high shadowing parameter considering the indoor environment $X_c\in [20, 70]$ and $\eta\in [6.0, 9.0]$. The recovering algorithms are applied to the validation set with various $\sigma^2\in [1.0, 5.0]$ and the best-performing $\sigma^2$ is selected. All the regularization parameter for BTD and DGM are set to be $10^{-4}$.
We set ${R} = 7$ and $L =4$ following \cite{shrestha2022deep} and \cite{zhang2020spectrum}, respectively. We do not include \texttt{KR} as it needs the real PSDs of the emitters, which are unavailable.

\subsubsection{DGM Training}
To train GAN for real data, we first generate samples that have large $\eta$ and small $X_c$ to simulate heavy shadowing conditions---as we know that the real data was collected from an indoor environment. We select $\eta$ ranging from 8 to 12 and $X_c$ ranging from 5 to 50. We simulate the SLF within a $14 \times 34 {\rm m}^2$ region. We then generate $10,000$ samples of such SLFs and train the GAN to learn the DGM. The GAN is trained using the \texttt{Adam} algorithm and a batch size of 256, for a maximum of 500 epochs. The initial learning rates of the discriminator and the generator are $2 \times 10^{-5}$  and $ 10^{-5}$, respectively. The GAN architecture is in Appendix \ref{Appendix:architecture}.

\begin{figure}[t!]
    \centering
    \includegraphics[width=\linewidth]{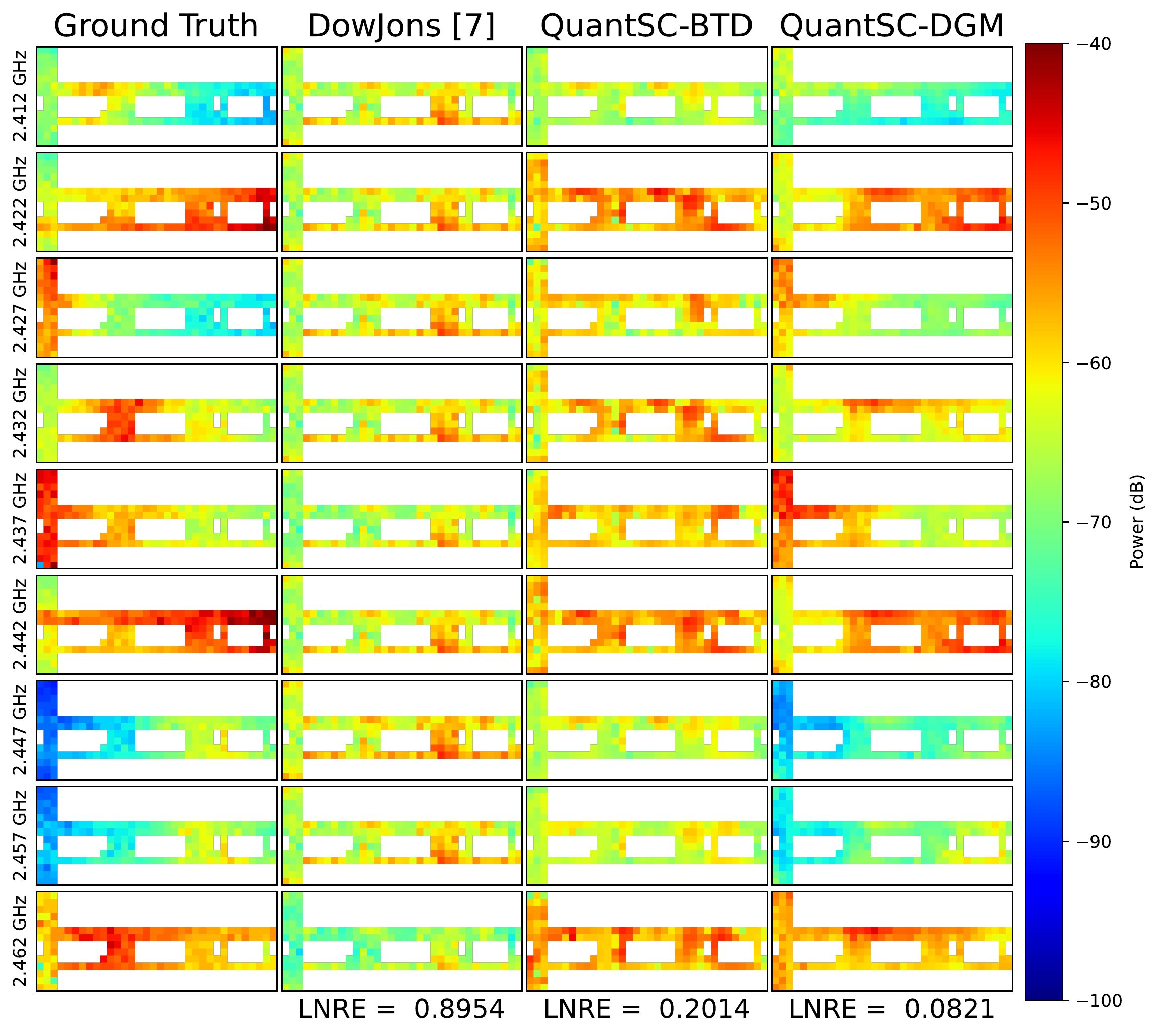}    \caption{Ground-truth and reconstructed radio maps on all the frequency bins;
    $\rho = 10
    \%$, $B=3$.}
    \label{fig:real_data}
\end{figure}

\subsubsection{Result}

Fig. \ref{fig:real_data} shows the reconstructed radio maps over all 9 frequencies by the methods under 3-bit quantization. 
One can see that the \texttt{DowJons} method is incapable of accurately estimating the true radio map in such a challenging scenario.
However, the proposed methods both output visually reasonable estimates, showing the effectiveness of our framework. As before, the estimated radio map by \texttt{QuantSC-DGM} is visually closer to the ground truth, but \texttt{QuantSC-BTD} offers satisfactory results without training a neural model off-line.

\begin{table}
\begin{center}
\caption{LNRE under various $\rho$'s for real data for $B=3$. }
\label{table:real_rho}
\resizebox{0.8\linewidth}{!}{
\begin{tabular}{| c | c | c | c |}
\hline
 $\rho$ & \texttt{DowJons}\cite{shrestha2022deep} & \texttt{QuantSC-BTD} & \texttt{QuantSC-DGM}\\
\hline
5\% & 0.9474 \tpm 0.1017  & 0.2527 \tpm 0.0110 & \textbf{ 0.1144 \tpm 0.0084} \\
\hline
10\% & 0.8944 \tpm 0.0790  &  0.2197 \tpm 0.0162 & \textbf{  0.0866 \tpm 0.0079 } \\
\hline
15\% & 0.8761 \tpm 0.0524  & 0.2061 \tpm 0.0126 & \textbf{ 0.0809 \tpm 0.0068 } \\
\hline
20\% & 0.8452 \tpm 0.0303  & 0.1993 \tpm 0.0098 & \textbf{ 0.0782 \tpm 0.0102 } \\
\hline
\end{tabular}}
\end{center}
\end{table}

Tables~\ref{table:real_rho}-\ref{table:real_bits} show the quantitative results under various $\rho$'s and $B$'s. The performance of the algorithms are consistent with what we saw in the simulations. Remarkably, we have no access to the environment parameters of the real data. Hence, it is likely that our generated simulated SLFs for training the DGM have nontrivial model mismatches. However, both the qualitative and quantitative evaluations suggest that the proposed method work reasonably well, which shows the robustness of the proposed approach.

\begin{table}
\begin{center}
\caption{LNRE under various bits for real data, $\rho =10\%$ }
\label{table:real_bits}
\resizebox{0.8\linewidth}{!}{

\begin{tabular}{| c | c | c | c |}
\hline
 Bits & \texttt{DowJons}\cite{shrestha2022deep} & \texttt{QuantSC-BTD} & \texttt{QuantSC-DGM}\\
\hline
1 & 0.9534 \tpm 0.0624 & 0.2842 \tpm 0.0162 & \textbf{ 0.0998 \tpm 0.0111 } \\
\hline
2 & 0.9264 \tpm 0.0233  & 0.2431 \tpm 0.0100 & \textbf{ 0.0897 \tpm 0.0076 } \\
\hline
3 & 0.8032 \tpm 0.0581 &  0.2045 \tpm 0.0095 & \textbf{  0.0879 \tpm 0.0082 } \\
\hline
4 & 0.7177 \tpm 0.0310 &  0.1934 \tpm 0.0128 & \textbf{  0.0862 \tpm 0.0073 } \\
\hline
5 & 0.5350 \tpm 0.0186 &  0.1887 \tpm 0.0101 & \textbf{  0.0812 \tpm 0.0092 } \\
\hline
6 & 0.2350 \tpm 0.0270 &  0.1810 \tpm 0.0137 & \textbf{  0.0797 \tpm  0.0057} \\
\hline
7 & 0.1809 \tpm 0.0148 &  0.1811 \tpm 0.0083 & \textbf{  0.0791 \tpm  0.0087} \\
\hline
8 & 0.1414 \tpm 0.0159 &  0.1803 \tpm 0.0070 & \textbf{  0.0786 \tpm 0.0072 } \\
\hline
\end{tabular}}
\end{center}
\end{table}

\section{Conclusion}
A novel framework has been introduced for SC that operates effectively with heavily quantized sensor measurements. Unlike previous approaches to provable SC, which assumed full-resolution real-valued measurements, the new maximum likelihood estimation (MLE)-based framework uses measurements discretized by a Gaussian quantizer. It is compatible with both BTD and DGM-based radio map representations, with the former being training-free, and the latter showing higher resilience to heavy shadowing. The recoverability of the framework has been characterized under realistic conditions, such as imperfect radio map modeling. Simulations and real-data experiments have demonstrated the effectiveness of the proposed approach. The proposed quantized SC framework offers a practical and realistic solution to radio map estimation, with provable guarantees.

\bibliographystyle{IEEEtran}
\bibliography{refs}

\appendices
\section{GAN Architecture}\label{Appendix:architecture}
Table \ref{table:arch} [left] presents the detailed architecture of the GAN used in our synthetic data experiments. The ``Deconv" term represents the block of operation stacked with transposed convolutional operations, followed by batch normalization and ReLU activations. Similarly, the ``Conv" term represents the block stacked with convolutional operations, followed by batch normalization and leaky ReLU activations. The ``Conv2d" refers to the convolutional operation followed by sigmoid activations. As mentioned, in order to ensure the nonnegativity of the generated SLFs, we use the sigmoid activation in the output layer of the generator. \#C, \#S, and \#P denote the number of channels, stride size, and pad size, respectively. Table \ref{table:arch} [right] presents the neural network structures used in the real data experiments.

\begin{table}[!h]
\caption{ GAN architecture in simulations. [Left] For synthetic data experiments. [Right] For real data experiments } \label{table:arch}
\begin{center}
\resizebox{0.48\linewidth}{!}{
\begin{tabular}{|c|c|c|c|c|}
\hline
\multicolumn{5}{|c|}{Generator} \\
\hline
 \textbf{Layer}& \textbf{Filter}& \textbf{\#C}& \textbf{\#S}&\textbf{\#P} \\
\hline
Deconv & $3 \times 3$ & 128 & 1 & 0\\
\hline
Deconv & $4 \times 4$ & 64 & 2 & 1\\
\hline
Deconv & $4 \times 4$ & 32 & 2 & 1\\
\hline
Deconv & $4 \times 4$ & 16 & 2 & 0\\
\hline
Deconv & $4 \times 4$ & 2 & 2 & 0\\
\hline
Conv2d & $4 \times 4$ & 1 & 1 & 0\\
\hline
\multicolumn{5}{|c|}{Discriminator} \\
\hline
Conv & $4 \times 4$ & 16 & 2 & 1\\
\hline
Conv & $4 \times 4$ & 32 & 2 & 1\\
\hline
Conv & $4 \times 4$ & 64 & 2 & 1\\
\hline
Conv & $4 \times 4$ & 128 & 2 & 1\\
\hline
Conv2d & $3 \times 3$ & 1 & 1 & 0\\
\hline
\end{tabular}
}
\resizebox{0.48\linewidth}{!}{
    
    \begin{tabular}{|c|c|c|c|c|}
    \hline
    \multicolumn{5}{|c|}{Generator} \\
    \hline
    \textbf{Layer} & \textbf{Filter}& \textbf{\#C}& \textbf{\#S}&\textbf{\#P} \\
    \hline
    Deconv & $3 \times 3$ & 128 & 1 & 0\\
    \hline
    Deconv & $3 \times 3$ & 64 & 2 & 1\\
    \hline
    Deconv & $3 \times 3$ & 32 & (1,2) & (0,1)\\
    \hline
    Deconv & $3 \times 3$ & 16 & 2 & 1\\
    \hline
    Deconv & $2 \times 2$ & 8 & (1,2) & 0\\
    \hline
    Deconv & $2 \times 2$ & 4 & 1 & 0\\
    \hline
    Conv2d & $2 \times 2$ & 1 & 1 & 0\\
    \hline
    \multicolumn{5}{|c|}{Discriminator} \\
    \hline
    Conv & $2 \times 2$ & 16 & 2 & 1\\
    \hline
    Conv & $4 \times 4$ & 32 & 1 & 1\\
    \hline
    Conv & $4 \times 4$ & 64 & 1 & (1,0)\\
    \hline
    Conv & $3 \times 3$ & 128 & 2 & 1\\
    \hline
    Conv & $3 \times 3$ & 256 & 2 & (1,0)\\
    \hline
    Conv2d & $2 \times 2$ & 1 & 1 & 0\\
    \hline
    \end{tabular}
}
\end{center}
\end{table}

\section{Proof of Lemma \ref{lemma:trans_function}}\label{proof:lemma1}
{
Let us assume that there are positive $\ell_{\min}$ and $\ell_{\max}$ such that the following holds:
\begin{align}\label{eq:lip_elementwise}
    \ell_{\rm min} &\leq \frac{| h(x) - h(y) |}{| x - y |} \leq \ell_{\rm max}\\
    \implies \ell_{\rm min} | x - y | &\leq | h(x) - h(y) | \leq \ell_{\rm max} | x - y |.
\end{align}
Applying the above to the coresponding entries of the two tensors $\tX$ and $\tX'$, and then squaring and summing over all the entries, we have
\begin{align}
    \ell^2_{\rm min} \| \tX - \tX' \|^2_{\rm F} &\leq  \| h(\tX) - h(\tX') \|^2_{\rm F}  \leq \ell^2_{\rm max} \| \tX - \tX' \|^2_{\rm F} \nonumber \\
   {\implies} \ell_{\rm min} &\leq  \frac{\| h(\tX) - h(\tX') \|_{\rm F}}{\| \tX - \tX' \|_{\rm F}}  \leq \ell_{\rm max}.
\end{align}
To find out $\ell_{\min}$ and $\ell_{\rm max}$, consider the following:
\begin{align}\label{eq:bound_min_max_log}
    & \min_{x,y \in [0, \alpha]} \frac{ | \log(x + a) - \log( y + a)| }{|x - y|} \nonumber\\
    & \leq \frac{\| h(\tX) - h(\tX') \|_{\rm F}}{\| \tX - \tX' \|_{\rm F}} \leq \max_{x,y \in [0,\alpha]} \frac{ | \log(x + a) - \log( y + a)| }{|x - y|} \nonumber\\
    & \stackrel{(b)}{\implies} \min_{x,y \in [0, \alpha]} \frac{ \log(x + a) - \log( y + a) }{|x - y|} \nonumber \\
    & \leq \frac{\| h(\tX) - h(\tX') \|_{\rm F}}{\| \tX - \tX' \|_{\rm F}} \leq \max_{x,y \in [0,\alpha]} \frac{  \log(x + a) - \log( y + a) }{x - y},
\end{align}
where $x$ and $y$ represent two arbitrary entries of $\tX$ and $\tX'$, respectively, (b) holds because the numerator and the denominator have the same sign.

Using the mean value theorem, there exists $z \in [x,y]$ such that
$$ \dot h(z) = \frac{  \log(x + a) - \log( y + a) }{x - y},$$
where $\dot h(z)$ is the derivative of $h$ at $z$. 
We know that
$\dot h(z) = \frac{1}{z+a}.$
Since $\dot h(z)$ is monotonically increasing with $z$, the minimum and maximum of \eqref{eq:bound_min_max_log} can be bounded by the minimum and maximum value of $\dot h(z)$. Hence,
\begin{align*}{}
    \min_{x,y \in [0,\alpha]} \frac{  \log(x + a) - \log( y + a) }{x - y} &= \min_{z \in [0,\alpha]} \frac{1}{z+a} = \frac{1}{\alpha + a}.
\end{align*}
Similarly, we have
    $\max_{x,y \in [0,\alpha]} \frac{  \log(x + a) - \log( y + a) }{x - y} = \frac{1}{a}.$
 Combining the above with \eqref{eq:bound_min_max_log} concludes the proof.

} 

\section{Proof of Fact \ref{fact:constants}}\label{proof:fact1}
 Note that $\tM$ is bounded if $\|\tX\|_{\infty}\leq \alpha$:
\begin{align*}
    \|\tM - \underline{\zero}\|_{\infty} &= \| h(\tX) - h((1-a)\underline{\one}) \|_{\infty}\\
    \implies \|\tM\|_{\infty} &\stackrel{(a)}{\leq} \max_{i,j,k} \frac{1}{a}| \tX(i,j,k) - (1-a) |\\
    &\stackrel{(b)}{\leq} \frac{1}{a} \max_{i,j,k} ( | \tX(i,j,k) | + |(1-a)|) \leq \frac{\alpha + |1-a|}{a},
\end{align*}
where $(a)$ used \eqref{eq:lip} and $(b)$ used the triangle inequality.

Note that $|\dot f_\ell (m)| = |\dot \Phi(b_\ell-m) - \dot \Phi(b_{\ell-1} - m)|  < \infty$ since $\dot \Phi(\cdot)$ is upper bounded for a given variance.
Moreover, the denominator
\begin{align*}
f_\ell (m) & = \Phi(b_\ell-m) - \Phi(b_{\ell-1} - m) \\
& = \int_{b_{\ell-1}}^{b_\ell} \frac{1}{\sqrt{2\pi} \sigma^2} \exp \left( \frac{-t^2}{\sigma^2}\right) dt > 0,   
\end{align*}
for given $b_\ell > b_{\ell-1}, \ell = \{2, \dots, Q\}$. Hence the supremum and infimum defined in \eqref{eq:constants} exist, provided that $\tX(i,j,k)$ is upper bounded by $\alpha$.

\section{Covering Numbers and Proof of Theorems} 
\subsection{Covering Numbers}\label{Appendix:covering}
We will use the following definition repeatedly:
\begin{Def}[Covering Number \cite{shalev2014understanding}] The covering number ${\sf N}(\mathcal{W}, \epsilon)$ of a set ${\cal W}$ with parameter $\epsilon>0$ is the smallest cardinality of any $\overline{\cal W}\subseteq {\cal W}$ such that for any $\w \in \cW$, there exists a $\overline{\w} \in \overline{\cW} \subset \cW$ satisfying
$\| \overline{\w} - \w\| \leq \epsilon$.
The discretized set $ \overline{\cW} $ is called the {\it $\epsilon$-net} of $\cW$. 
\end{Def}
Roughly speaking, the continuous set $\cX$ can be approximated by a discrete set whose cardinality is $  {\sf N}\left( \cX, \epsilon \right)$.

 \begin{Lemma}[Covering Number of ${\cal X}^{\rm BTD}$]\label{lemma:coverno_btd}
 The covering number of ${\cal X}^{\rm BTD}$, i.e., ${\sf N}(\cX^{\rm BTD}, \epsilon)$, is upper bounded by
\begin{align}\label{eq:coverBTD}
 \bigg( \frac{3 (\kappa + \beta) R}{\epsilon} \bigg)^{((I+J)L+K)R} (\beta)^{(I+J)LR}  \bigg(\frac{\kappa}{2} \bigg)^{RK}. 
\end{align} 
\end{Lemma}

\begin{IEEEproof}
Let $\overline{\cA}_r$ denote an $\frac{\epsilon}{2\sqrt{\beta}}$-net of $\cA_r = \{\A_r  \in \bbR^{I \times L} ~ | ~ \|\A_r\|_{\rm F} \leq  \sqrt{\beta}  , \A_r \geq 0\}$. Similarly, let $\overline{\cB}_r$ denote an $\frac{\epsilon}{2\sqrt{\beta}}$-net of $\cB_r = \{\B_r \in \bbR^{J \times L} ~ | ~  \|\B_r\|_{\rm F} \leq  \sqrt{\beta}, \B_r \geq 0  \}$.
According to \cite[Lemma 4.1]{pollard1990empirical}, the covering number of an $\epsilon$-net of $D$-dimensional Euclidean ball, i.e., $\{\x \in \bbR^D | \| \x\|_2 \leq \delta\}$ is upper bounded by $ \left (\frac{3 \delta}{\epsilon} \right)^{D}$.
Therefore, 
\begin{align}
 |\overline{\cal A}_r|\leq  \bigg( \frac{6 \beta }{\epsilon} \bigg)^{IL} \frac{1}{2^{IL}},~
 |\overline{\cal B}_r|\leq  \bigg( \frac{6 \beta }{\epsilon} \bigg)^{JL} \frac{1}{2^{JL}},
\end{align}
where the constant factors $\frac{1}{2^{IL}}$ and $\frac{1}{2^{JL}}$ are due to non-negativity constraints in the sets $\cA_r$ and $\cB_r$, respectively.
Next, consider the set ${\cal F}_r = \{ \S_r = \A_r \B_r ^\top \in \bbR^{I \times J} | {\rm rank}(\S_r) \leq L, \| \S_r \|_{\rm F} \leq \beta, \S_r \geq 0, \A_r \in \cA_r, \B_r \in \cB_r \}$. We can construct an $\epsilon$-net of ${\cal F}_r$ such that the following holds:
\begin{subequations}
\begin{align}
    \| & \S_r - \overline{\S}_r \|_{\rm F} = \|\A_r \B_r^\top - \overline{\A}_r \overline{\B}^\top_r \|_{\rm F} \nonumber \\
    &\leq \|\A_r (\B_r - \overline{\B}_r)^\top \|_{\rm F} +  \|(\A_r - \overline{\A}_r) \overline{\B}_r^\top \|_{\rm F} \label{eq:traingles_AB} \\
    &\leq \|\A_r \|_{\rm F} \|\B_r^\top - \overline{\B}_r^\top \|_{\rm F} +  \|\A_r - \overline{\A}_r\|_{\rm F} \|\overline{\B}_r^\top \|_{\rm F} \label{eq:cauchy_AB} \\
    &\leq \sqrt{\beta} \left (\frac{\epsilon}{2\sqrt{\beta}} + \frac{\epsilon}{2\sqrt{\beta}} \right ) \leq \epsilon, \nonumber
\end{align}
\end{subequations}
where \eqref{eq:traingles_AB} uses the triangle inequality and \eqref{eq:cauchy_AB} follows from the Cauchy-Schwarz inequality. The cardinality of an $\epsilon-$net of ${\cal F}_r$ is equal to the product of cardinalities of the sets $\overline{\cal A}_r$ and $\overline{\cal B}_r$. Let $\overline{\cal F}_r$ be the $\frac{\epsilon}{R(\kappa+\beta)}$-net of ${\cal F}_r$, then
\begin{align}
    |\overline{\cal F}_r|\leq  \bigg( \nicefrac{3\beta R (\kappa + \beta)}{\epsilon} \bigg)^{(I+J)L}.
\end{align}
Now, let $\overline{\cal C}_r$ denote an $\frac{\epsilon}{R(\kappa+\beta)}$-net of ${\cal C}_r=\{\bc_r \in \bbR^{K} | \|\bc_r\|_2\leq \kappa, \bc_r \geq 0 \}$. Then, using \cite[Lemma 4.1]{pollard1990empirical}, we get
\begin{align}
|\overline{\cal C}_r|\leq \bigg( \nicefrac{3\kappa R (\kappa + \beta)}{\epsilon} \bigg)^{K} \frac{1}{2^K}.
\end{align}

Finally, consider a discrete set $\overline{\cal X}^{\rm BTD}$. We hope to construct this set such that for every $\tX\in {\cal X}^{\rm BTD}$, there exists an
$\overline{\tX} \in \overline{\cal X}^{\rm BTD}$ such that $\|\tX -\overline{\tX} \|_{\rm F}\leq \epsilon$. To this end, we let 
\begin{align*}
    \overline{\cal X}^{\rm BTD}=\left\{\overline{\tX}~|~\overline{\tX}=\sum_{\r=1}^R \overline{\S}_r\circ{\bm c}_r,~\overline{\S}_r\in \overline{\cal F}_r,\overline{\bm c}_r\in \overline{\cal C}_r \right\}.
\end{align*}
To see that the above construction is an $\epsilon$-net of ${\cal X}^{\rm BTD}$, 
we show that there exists an $\overline{\tX}\in \overline{\cal X}^{\rm BTD}$ that satisfies the following chain of inequalities for any $\tX\in {\cal X}^{\rm BTD}$:
\begin{subequations}
    \begin{align}
     &\|\tX -\overline{\tX} \|_{\rm F} =\left\|   \sum_{r=1}^R \S_r\circ \bm c_r - \sum_{\r=1}^R \overline{\S}_r\circ \overline{\bm c}_r  \right\|_{\rm F} \nonumber\\
     & \stackrel{(a)}{=} \|\bm C\S - \overline{\bm C}\overline{\S}\|_{\rm F}  \leq \| \bm C(\bm S - \overline{\bm S})        \|_{\rm F} + \|(\bm C-\overline{\bm C})\overline{\bm S}\|_{\rm F}  \nonumber \\
         &\leq \| {\C} \|_{\rm F} \| {\bm S} -  \overline{\bm S} \|_{\rm F} + \|{\C} -  \overline{\C} \|_{\rm F} \|\overline{\bm S}\|_{\rm F} \nonumber \\
         & =\sqrt{R}\kappa \left(  \sum_{r=1}^R\|\S_r - \overline{\S}_r\|_{\rm F}^2         \right)^\frac{1}{2} + \sqrt{R} \beta \left(  \sum_{r=1}^R\|\bm c_r - \overline{\bm c}_r\|_{2}^2         \right)^\frac{1}{2} \nonumber\\
         &\stackrel{(b)}{\leq} \sqrt{R} \kappa \left ( R \left( \frac{\epsilon}{R(\kappa + \beta)} \right)^2 \right)^\frac{1}{2} + \sqrt{R} \beta \left ( R \left( \frac{\epsilon}{R(\kappa+ \beta)} \right)^2 \right)^\frac{1}{2} \nonumber \\
         &\leq \epsilon \nonumber
    \end{align}
\end{subequations}
where $\bm C=[\bm c_1,\ldots,\bm c_R]$ and $\bm S= [\bm s_1,\ldots,\bm s_R]^\top$ with $\bm s_r={\rm vec}(\bm S_r)$ for $r\in[R]$ and $(a)$ is based on the matrix representation of the outer product; see \cite{shrestha2022deep} \cite{ding2023fast}.
The inequality $(b)$ holds by picking the $\overline{\S}_r$ from $\overline{\cal F}_r$ that has the smallest distance to $\bm S_r$.
The same applies to the pick of $\overline{\bm c}_r$.

The cardinality of $\overline{\cal X}^{\rm BTD}$ is equal to the product of the cardinality of $R$ sets of $\overline{\cal F}_r$ and $\overline{\cal C}_r$, respectively. Hence, the covering number of set $\cX^{\rm BTD}$ is upper bounded by \eqref{eq:coverBTD}.
\end{IEEEproof}

Under Assumption~\ref{ass:generator_structure}, the covering number of $\cX^{\rm DGM}$ is bounded as well \cite{shrestha2022deep}:
\begin{Fact}[Covering Number of ${\cal X}^{\rm DGM}$ \cite{shrestha2022deep}]\label{fact:cover}
Under Assumption~\ref{ass:generator_structure},
the covering number of $\cX^{\rm DGM}$ satisfies 
\begin{align}\label{eq:cover}
   {\sf N}\left( \cX^{\rm DGM},\epsilon \right)  \leq \left( \nicefrac{3R(\beta + \kappa)}{\epsilon} \right ) ^{R(K+D)} ( \nicefrac{\kappa}{2} )^{RK} (Pq)^{RD}.
\end{align}
\end{Fact}

\subsection{Proof of Theorems \ref{thm:btd} and \ref{thm:dgm} } \label{Appendix:thm_proof}
Let us define $\cT = \{ (0,0),\dots,(I,J) \}$. We take $N$ samples denoted by $\omega_1, \dots, \omega_N$ from $\cT$ uniformly with replacement and collect the indices in $\bOmega = \{ \omega_1, \omega_2\dots,\omega_N \}$. 
Then, for every $\omega_i \in \bOmega, k \in [K]$, $\tY_{\omega_i k}$ is observed with the probability mass function defined by $\{f_\ell (\tM_{\omega_i k})\}_{\ell=1}^Q$. 
Here, we define a shorthand notation by letting $\tY_{\omega k} = \tY(i,j,k)$ and $\tY_{\omega :} = \tY(i,j,:)$, where $\omega \in \bOmega$. We use the similar notation for $\tM$.
For any two fibers, $\tM_{\omega :}$ and $\tY_{\omega :}, ~ \forall \omega \in \bOmega$ the function $g: \bbR^K \times \bbR^K \to \bbR$ is defined as follows:
\begin{align}
    g(\tM_{\omega :}, \tY_{\omega :}) = \sum_{k=1}^K \sum_{\ell=1}^{Q} \mathbb{1}_{[\tY_{ijk}=\ell]} \log \left (  \frac{1}{ f_{\ell}(\tM_{\omega k}) }  \right).
\end{align}
Recall that $\tM = h(\tX)$. Then we define
\begin{align*}
    F_{\bOmega, \tY}(\tM) &= \frac{1}{N} \sum_{(i,j) \in \bOmega} g(\tM(i,j,:),  \tY(i,j,:)) \\
    & = \frac{1}{N} \sum_{(i,j) \in \bOmega} \sum_{k=1}^K \sum_{\ell=1}^{Q} \mathbb{1}_{[\tY_{ijk}=\ell]} \log \left (  \frac{1}{ f_{\ell}(\tM_{ijk}) }  \right).
\end{align*}

Let, $\gtM = h(\gtX)$ be the ground truth tensor and $\otM = h(\tX^\star)$ be the optimal solution of the optimization problem \eqref{eq:model_qsc}.
\begin{align}
    &\bbE_{\omega, \tY} [g(\gtM_{\omega :},  \tY_{\omega :}) - g(\tM^\star_{\omega : },  \tY_{\omega :})] \nonumber\\
    & = \sum_{(i,j) \in \cT} \frac{1}{IJ} \bbE_{\tY | \omega = (i,j)} \left (g(\gtM_{\omega :}, \tY_{\omega :}) - g(\otM_{\omega :}, \tY_{\omega :})  \right ) \nonumber \\
    & = \frac{1}{IJ} \left ( \sum_{(i,j) \in \cT} \sum_{k=1}^K \sum_{\ell=1}^{Q}  P(\Y_{\omega k} = \ell) \log \left (  \frac{ f_{\ell}(\otM_{\omega k}) }{ f_{\ell}(\gtM_{\omega k}) } \right) \right ) \nonumber \\
    & =  \frac{K}{IJK} \sum_{(i,j,k) \in [I] \times [J] \times [K]} \sum_{\ell=1}^{Q} f_{\ell}(\gtM_{\omega k}) \log \left (  \frac{ f_{\ell}(\otM_{\omega k}) }{ f_{\ell}(\gtM_{\omega k}) } \right) \nonumber \\
    &= -K \cdot \text{KL}(\otM||\gtM),
    \label{eq:KL_excess_risk}
\end{align}
where ${\rm KL}(\tM^\star || \tM^\natural) = \frac{1}{IJK} \sum_{i,j,k}  {\rm KL}(\tM^\star_{ijk} || \tM^\natural_{ijk}) $ and ${\rm KL}(\tM^\star_{ijk} || \tM^\natural_{ijk}) = \sum_{\ell=1}^Q f_{\ell}(\tM^\natural_{ijk}) \log \frac{f_{\ell} (\tM^\natural_{ijk})}{f_\ell(\tM^\star_{ijk})}$, which follow the definitions in \cite{ghadermarzy2018learning}.

Let us denote $\widetilde{\tM} = \argmin{\tM \in \cM}  \bbE_{\omega, \tY}[g(\tM_{\omega :}, \tY_{\omega: }) ]$,
where ${\cal M}\in \{{\cal M}^{\rm BTD},{\cal M}^{\rm DGM} \}$, depending on the model that we use.
That is, $\widetilde{\tM}$ is the solution that we can obtain from the expected version of our MLE.

Also, we have $F_{\bOmega, \tY}(\otM) \leq F_{\bOmega, \tY}(\wtM)$, which implies the following chain of inequalities:
\begin{align}
    &0 \leq F_{\bOmega, \tY}(\wtM) -F_{\bOmega, \tY}(\gtM) + F_{\bOmega, \tY}(\gtM) - F_{\bOmega, \tY}(\tM^\star) \nonumber \\
    &\leq \bbE_{\omega, \tY}[g(\gtM_{\omega :}, \tY_{\omega :}) ] - \bbE_{\omega, \tY}[g(\otM_{\omega :}, \tY_{\omega :}) ] + F_{\bOmega, \tY}(\gtM) \nonumber\\
    &- \bbE_{\omega, \tY}[g(\gtM_{\omega :}, \tY_{\omega :}) ] + \bbE_{\omega, \tY}[g(\otM_{\omega :}, \tY_{\omega :}) ] - F_{\bOmega, \tY}(\tM^\star)  \nonumber\\
    &+ F_{\bOmega, \tY}(\wtM) - F_{\bOmega, \tY}(\gtM) 
\end{align}
\begin{align}
    & \stackrel{(a)}{\leq} \bbE_{\omega, \tY}[g(\gtM_{\omega :}, \tY_{\omega :}) - g(\otM_{\omega :}, \tY_{\omega :}) ] \nonumber\\
    &+ |F_{\bOmega, \tY}(\gtM) - \bbE_{\omega, \tY}[g(\gtM_{\omega :}, \tY_{\omega :}) ] | \nonumber\\
    &+ |F_{\bOmega, \tY}(\otM) - \bbE_{\omega, \tY}[g(\otM_{\omega :}, \tY_{\omega :}) ]|  + |F_{\bOmega, \tY}(\widetilde{\tM}) - F_{\bOmega, \tY}(\gtM)| \nonumber \\
    & \stackrel{(b)}{\implies} K \cdot \text{KL}(\otM||\gtM)  \leq  |F_{\bOmega, \tY}(\widetilde{\tM}) - F_{\bOmega, \tY}(\gtM)| \nonumber\\
    & + | F_{\bOmega, \tY}(\gtM) - \bbE_{\omega, \tY}[g(\gtM_{\omega : }, \tY_{\omega : }) ]| \nonumber \\
    & + \sup_{\tM \in \cM} | F_{\bOmega, \tY}(\tM) - \bbE_{\omega, \tY}[g(\tM_{\omega :}, \tY_{\omega :}) ]|, \label{eq:bounding_term}
\end{align}
where $(a)$ is by the triangle inequality and $(b)$ follows  \eqref{eq:KL_excess_risk}.

Recall that $\min_{\widetilde{\tX}\in \cX_{R, \g_{\btheta}}}\|\widetilde{\tX} - \gtX \|_{\infty} \leq \nu$, where $\nu \in \{\nu^{\rm BTD}, \nu^{\rm DGM} \}$ and $\nu \geq 0$ is a constant. The first term on the R.H.S. of \eqref{eq:bounding_term} can be bounded as follows
\begin{align}
    &|F_{\bOmega, \tY}(\widetilde{\tM}) - F_{\bOmega, \tY}(\gtM)| \label{eq:represent_bound} \\
    & = \bigg | \frac{1}{N} \sum_{(i,j) \in \bOmega} \sum_{k=1}^K \sum_{\ell=1}^{Q} \mathbb{1}_{[\tY_{ijk}=\ell]} \left( \log ( f_{\ell}(\gtM_{ijk}))- \log (f_{\ell}(\widetilde{\tM}_{ijk}) ) \right) \bigg | \nonumber \\
    & \stackrel{(a)}{\leq}  \frac{1}{N} \sum_{(i,j) \in \bOmega} \sum_{k=1}^K \sum_{\ell=1}^{Q} \mathbb{1}_{[\tY_{ijk}=\ell]} \left | \log (  f_{\ell}(\gtM_{ijk})  ) - \log (  f_{\ell}(\widetilde{\tM}_{ijk})  ) \right | \nonumber \\
    & \stackrel{(b)}{\leq}  \frac{1}{N} \sum_{(i,j) \in \bOmega} \sum_{k=1}^K \sum_{\ell=1}^{Q} \mathbb{1}_{[\tY_{ijk}=\ell]} \lalpha \frac{1}{a} \left \| \gtX - \widetilde{\tX} \right \|_{\infty} \nonumber\\
    & \stackrel{(c)}{\leq}  \frac{\lalpha \nu}{a N} \sum_{(i,j) \in \bOmega} \sum_{k=1}^K \sum_{\ell=1}^{Q} \mathbb{1}_{[\tY_{ijk}=\ell]} \leq  \frac{\lalpha \nu}{a N} \sum_{(i,j) \in \bOmega} \sum_{k=1}^K 1 \leq \frac{K \lalpha \nu}{a} \nonumber. 
\end{align}
In the above, $(a)$ used the triangle inequality. In addition, $(b)$ used the property of log-concavity, Fact \ref{fact:constants}, and \eqref{eq:lip_elementwise}. The $(c)$ part used the fact that $\sum_{\ell=1}^{Q} \mathbb{1}_{[ \tY_{ijk} = \ell] } = 1$.

The second term in \eqref{eq:bounding_term} can be bounded using Hoeffding's inequality \cite{hoeffding1963probability} on independent random variable $Z_{\omega} = \bm g(\tM_{\omega: }, \tY_{\omega:})$ and $Z_\omega \in [0, K\ualpha]$. Then with probability greater than $1- \delta$, the following holds:
\begin{align}\label{eq:hoeffding_bound}
| F_{\bOmega, \tY}(\gtM) - \bbE_{\omega, \tY}[g(\gtM_{\omega:}, \tY_{\omega:}) ]| \leq \sqrt{\nicefrac{(K \ualpha)^2\log(\frac{1}{\delta})}{2N}}.
\end{align}

For the third term in \eqref{eq:bounding_term}, observe that $F_{\bOmega, \tY}(\M)$ is a sample average of the samples of joint random variable $(\omega, \tY)$ drawn independently from their joint distribution. And the corresponding expected value is $\bbE_{\omega, \tY}[g(\tM_{\omega:}, \tY_{\omega:})]$. Hence, the third term can be bounded via Rademacher Complexity-based arguments \cite{shalev2014understanding}. Specifically, with probability at least $1-\delta$, the following holds
\begin{align}\label{eq:generalization_error_bound}
    &\sup_{\tM \in \cM} | F_{\bOmega, \tY}(\tM) - \bbE_{\omega, \tY}[g(\tM_{\omega :}, \tY_{\omega :}) ]| \nonumber \\
    &\leq 2\widehat{\cR}(\cG) + K \ualpha \sqrt{ \frac{8\log(\frac{2}{\delta})}{N} },
\end{align}
where $\widehat{\cR}(\cG)$ is the empirical Rademacher complexity of the function class $\cG$ defined as follows:
$$ \cG = \{ (\omega, \tY) \mapsto g(\tM_{\omega :}, \tY_{\omega :}) | \tM = h(\tX), \tX \in \cX \}.$$
Using Talagrand Lemma \cite[Lemma 4.2]{mohri2018foundations} along with Lemma \ref{lemma:trans_function} and Lipschitz property of $g(\tM_{\omega :}, \tY_{\omega :})$ in terms of $f_{\ell}$ in Fact \ref{fact:constants},
$$ \widehat{\cR}(\cG) \leq \frac{\lalpha}{a} \widehat{\cR}(\cX),$$
where ${\cal X}\in \{{\cal X}^{\rm BTD},{\cal X}^{\rm DGM} \}$. Using Dudley's entropy integral \cite[Lemma A.5]{bartlett2017spectrally}, one can bound the empirical Rademacher complexity  by the covering number (cf. Appendix~\ref{Appendix:covering}) of the set as follows:
\begin{align}\label{eq:rademacher_bound}
    \widehat{\cR}(\cX ) \leq \zeta(\cX, \mu),
\end{align}
where \begin{align}\label{eq:zeta_definition}
\zeta(\cX, \mu) = \inf_{\mu>0} \left ( \frac{4\mu}{\sqrt{N}} + \frac{12}{N} \int_{\mu}^{ \sqrt{N} } \sqrt{\log {\sf N}(\cX , \epsilon)} d\epsilon \right).
\end{align}

Combining \eqref{eq:bounding_term}, \eqref{eq:represent_bound}, \eqref{eq:hoeffding_bound}, \eqref{eq:generalization_error_bound}, and \eqref{eq:rademacher_bound}, we get
\begin{align}\label{eq:kl_bound}
    &\text{KL}(\M^\star||\gtM) \leq \\
    &\frac{2 \lalpha \zeta(\cX, \mu)}{a K} + \frac{K \ualpha}{K} \sqrt{\frac{\log(\frac{1}{\delta})}{2N}} + \frac{\ualpha K}{K}  \sqrt{ \frac{8\log(\frac{2}{\delta})}{N} }  + \frac{K \lalpha \nu}{a K} \nonumber.
\end{align}

The relation between Hellinger distance and KL divergence \cite{davenport20141} is as follows:
\begin{align}
    d_H^2(\otM, \gtM) \leq \text{KL}(\otM||\gtM).
    \label{eq:hellinger_kl}
\end{align}

From \cite[Lemma 2]{cao2015categorical} and Fact \ref{fact:constants},  we can bound Hellinger distance using the following:
\begin{align}
    \frac{\falpha}{4} \frac{\| \otM - \gtM \|_{\rm F}^2}{IJK} \leq d_H^2(\otM, \gtM).
    \label{eq:mse_hellinger}
\end{align}

Combining \eqref{eq:hellinger_kl}, \eqref{eq:mse_hellinger} and Lemma \ref{lemma:trans_function}, we get 
\begin{align}\label{eq:kl_frob_relation}
    \frac{\| \otM - \gtM  \|_{\rm F}^2}{IJK} & \leq \frac{4}{\falpha} \text{KL}(\otM||\gtM) \nonumber \\
    \implies \frac{\| \tX^\star - \gtX  \|_{\rm F}^2}{IJK} &\leq \frac{4(\alpha + a)^2}{\falpha} \text{KL}(\otM||\gtM).
\end{align}

Let us define the following constants,
\begin{align}\label{eq:C1}
    C_1 = \frac{4(\alpha + a)^2}{\falpha}, \text{ and } C_2 = \frac{\lalpha}{a}.
\end{align}

Combining \eqref{eq:kl_bound}, \eqref{eq:C1} and \eqref{eq:kl_frob_relation},
\begin{align}\label{eq:raw_bound}
    &\frac{\| \tX^\star - \gtX \|_{\rm F}^2}{IJK} \leq \\
    &C_1 \bigg ( \frac{2 C_2 \zeta(\cX, \mu)}{K}  + \ualpha \sqrt{\frac{\log(\frac{1}{\delta})}{2N}}  +  \ualpha \sqrt{ \frac{8\log(\frac{2}{\delta})}{N} } + C_2 \nu \bigg ). \nonumber
\end{align}
Since $( \log {\sf N}(\cX , \epsilon) )^{1/2}$ increases with the decrease in $\epsilon$, the integral in \eqref{eq:zeta_definition}
can be upper bounded as,
\begin{align}
    &\int_{\mu}^{ \sqrt{N} }  ( \log {\sf N}(\cX , \epsilon) )^{1/2} d\epsilon &\leq \sqrt{N} ( \log {\sf N}(\cX , \mu) )^{1/2}.
    \label{eq:integral_bound}
\end{align}

In the next two subsections, we will prove Theorems~\ref{thm:btd} and \ref{thm:dgm}, respectively, using the above technical preparations.

\subsubsection{Proof of Theorem \ref{thm:btd} }
Setting $\mu=\sqrt{R}$ and $\cX = \cX^{\rm BTD}$ in $\zeta(\cX, \mu)$, and using Lemma \ref{lemma:coverno_btd} and \eqref{eq:integral_bound},
 \begin{align}
     &\int_{\mu}^{ \sqrt{N} }  ( \log {\sf N}(\cX^{\rm BTD} , \epsilon) )^{1/2} d\epsilon 
     \nonumber \\
    &= \sqrt{NR} \bigg( \sqrt{((I+J)L+K) \log ( 3\sqrt{R}(\beta + \kappa) )} \nonumber \\
    & + \sqrt{K\log(\kappa/2)}+ {(I+J)L}\log(\beta) \bigg ) = \frac{\sqrt{NR}}{3} \tau, \label{eq:integration_btd}
\end{align}
where we have
 \begin{align*} \tau &= 3 \bigg( \sqrt{((I+J)L+K) \log ( 3\sqrt{R}(\beta + \kappa) )}\\
& + \sqrt{ K\log(\kappa/2)}+ {(I+J)L}\log(\beta) \bigg ). \end{align*}
Combining \eqref{eq:integration_btd} and \eqref{eq:zeta_definition},
\begin{align}\label{eq:zeta_btd}
    \zeta(\cX^{\rm BTD}, \sqrt{R}) = 4(1+\tau)\sqrt{\frac{R}{N}}.
\end{align} 
Combining \eqref{eq:raw_bound} and \eqref{eq:zeta_btd},
\begin{align}\label{eq:overal_btd}
    \frac{\| \tX^\star - \gtX \|_{\rm F}^2}{IJK} \leq&  \frac{8 C_1 C_2 (1+\tau)}{K}\sqrt{\frac{R}{N}} + \ualpha C_1 \sqrt{\frac{\log(\frac{1}{\delta})}{2N}} \nonumber \\
    &  +  \ualpha C_1 \sqrt{ \frac{8\log(\frac{2}{\delta})}{N} } +  C_1 C_2\nu^{\rm BTD}.
\end{align}
Note that as both the second and third terms in \eqref{eq:bounding_term} are upper bounded with probability at least $1-\delta$, the over all upper bound in \eqref{eq:overal_btd} holds with probability at least $1-2\delta$. This completes the proof of Theorem \ref{thm:btd}.

\subsubsection{Proof of Theorem \ref{thm:dgm}}
Setting $\mu=\sqrt{R}$ and $\cX = \cX^{\rm DGM}$ in $\zeta(\cX, \mu)$, and using Fact \ref{fact:cover} and \eqref{eq:integral_bound},
\begin{align}
     &\int_{\mu}^{ \sqrt{N} }  ( \log {\sf N}(\cX^{\rm DGM} , \epsilon) )^{1/2} d\epsilon \nonumber \\
    &= \sqrt{NR} \sqrt{(K+D) \log ( 3\sqrt{R}(\beta + \kappa) )+ K\log(\kappa/2)}+ D\log(Pq) \nonumber \\
    &= \sqrt{NR} \sqrt{K \log ( \frac{3}{2}\sqrt{R}\kappa(\beta + \kappa) )+ D\log ( 3\sqrt{R}Pq(\beta + \kappa) ) } \nonumber \\
    &= \frac{\sqrt{NR}}{3} \tau, \label{eq:integration_dgm}
\end{align}
where we have $$\tau = 3\sqrt{K \log ( \frac{3}{2}}\sqrt{R}\kappa(\beta + \kappa) )+ D\log ( 3\sqrt{R}Pq(\beta + \kappa) ).$$
Combining \eqref{eq:integration_dgm} and \eqref{eq:zeta_definition},
\begin{align}\label{eq:zeta_dgm}
  \zeta(\cX^{\rm DGM}, \sqrt{R}) = 4(1+\tau)\sqrt{\frac{R}{N}}.
\end{align}
Combining \eqref{eq:raw_bound} and \eqref{eq:zeta_dgm}, we have the following holds with probablity at least $1-2\delta$:
\begin{align}
    \frac{\| \tX^\star - \gtX \|_{\rm F}^2}{IJK} \leq&  \frac{8 C_1 C_2(1+\tau)}{K}\sqrt{\frac{R}{N}} + \ualpha C_1 \sqrt{\frac{\log(\frac{1}{\delta})}{2N}} \nonumber \\
    &  +  \ualpha C_1 \sqrt{ \frac{8\log(\frac{2}{\delta})}{N} } + C_1 C_2 \nu^{\rm DGM}.
\end{align}

This completes the proof of Theorem \ref{thm:dgm}.

\end{document}